\documentclass[a4paper,11pt]{article}
\usepackage{jheppub}[=2022-01-01]
\usepackage[T1]{fontenc}
\usepackage{lmodern}

\usepackage{amstext}
\usepackage{amsfonts}
\usepackage{mathrsfs}
\usepackage{amsmath}
\usepackage{amssymb}
\usepackage{diagbox}
\usepackage{subfig}
\usepackage{bm}
\usepackage{multirow}
\usepackage{cancel}
\usepackage{extarrows}
\usepackage{tensor}
\usepackage{tikz}
\usepackage{xcolor}
\usepackage{graphicx}
\usepackage[low-sup]{subdepth}
\usepackage{caption}
\usepackage{subcaption}
\usepackage[linktocpage=true]{hyperref}
\usepackage{here}
\allowdisplaybreaks[4]
\usepackage{indentfirst}
\usepackage{isodateo}
\usepackage{enumerate}

\newcommand{\beq}{\begin{equation}}
	\newcommand{\eeq}{\end{equation}}
\newcommand{\ba}{\begin{array}}
	\newcommand{\ea}{\end{array}}
\newcommand{\beqa}{\begin{eqnarray}}
	\newcommand{\eeqa}{\end{eqnarray}}
\newcommand{\beqs}{\begin{subequations}}
	\newcommand{\eeqs}{\end{subequations}}

\renewcommand{\rm}{\mathrm}

\def\leqq{\leqslant}
\def\geqq{\geqslant}
\def\({\left(}
\def\){\right)}
\def\[{\left[}
\def\]{\right]}

\def\nn{\nonumber}

\def\ito{\!\rightarrow\!}

\makeatletter
\newsavebox\myboxA
\newsavebox\myboxB
\newlength\mylenA
\newcommand*\xoverline[2][0.63]{%
\sbox{\myboxA}{$\m@th#2$}%
\setbox\myboxB\null
\ht\myboxB=\ht\myboxA%
\dp\myboxB=\dp\myboxA%
\wd\myboxB=#1\wd\myboxA
\sbox\myboxB{$\m@th\overline{\copy\myboxB}$}
	\setlength\mylenA{\the\wd\myboxA}
	\addtolength\mylenA{-\the\wd\myboxB}%
	\ifdim\wd\myboxB<\wd\myboxA%
	\rlap{\hskip 1.2\mylenA\usebox\myboxB}{\usebox\myboxA}%
	\else
	\hskip -0.5\mylenA\rlap{\usebox\myboxA}{\hskip 0.5\mylenA\usebox\myboxB}%
	\fi}

\def\ba{\bar{a}}

\def\A{\mathcal{A}}

\def\gh{\hat{g}}

\def\ii{\text{i}}

\def\bz{\bar{z}}

\def\si{\sigma}

\def\nn{\nonumber}
\def\vs{\vspace*{1mm}}

\def\hs{\hspace*{0.3mm}}

\def\hsm{\hspace*{-0.3mm}}

\def\ii{\rm{i}}
\def\di{\rm{d}}

\def\A{\mathcal{A}}

\def\bz{\bar{z}}
\def\bp{\bar{\partial}}
\def\End{\end{document}}

\title{%
Internal-loop-state decomposition and double copy of one-loop string integrands 
with applications to beta functions
}

\author[a,b]{Yao Li 
}
\emailAdd{neolee@mit.edu
}

\affiliation[a]{Center for Theoretical Physics, Massachusetts Institute of Technology,Cambridge, MA 02139, USA}

\affiliation[b]{
Shanghai Key Laboratory for Particle Physics and Cosmology,
Shanghai Jiao Tong University, Shanghai, China}

\abstract{\\ 
Following the ideas of Refs.~\cite{Tourkine:2012vx,Geyer:2015jch}, we systematically reformulate and generalize the decomposition of one-loop string integrands according to the internal states propagating in a chosen loop channel, extending the construction to both bosonic and heterotic string theories. Using the chiral-splitting formalism, we make manifest a double-copy structure between the left- and right-moving internal loop states. As an application, we analyze the one-loop beta functions of gauge and gravitational couplings from the low-energy field-theory limit of heterotic one-loop three-point amplitudes under a naive $T^6$ compactification. Although maximal supersymmetry forces these beta functions to vanish, our internal-state decomposition provides model-independent expressions for the contributions of different internal loop states. We further show explicitly that both the gravitational beta function and the gravitational correction to the gauge beta function vanish.  
}

\begin{document}
\maketitle

\section{\hspace*{-2.5mm}Introduction}
\label{sec:introduction}
The double-copy relation reveals a profound connection between perturbative scattering amplitudes in gauge theories and gravitational theories, expressed symbolically as $\text{GR} = \text{YM}^2$. A typical $L$-loop $n$-point scattering amplitude with a manifest double-copy structure can be written as 
\begin{equation}\label{eq:double-copy}
	A^n_L\sim \int d^{LD} l \sum_i \frac{1}{S_i}\frac{N_i \tilde{N}_i}{D_i}, 
\end{equation}
where $l$ denotes the loop momentum, $i$ runs over all possible $\phi^3$ Feynman diagrams, $D_i$ denotes the product of the corresponding scalar propagators, $S_i$ denotes possible symmetry factors, and $N_i$ and $\tilde{N}_i$ are BCJ numerators satisfying color-kinematics duality \cite{Bern:2008qj,Bern:2019prr}.   

Historically, the tree-level double copy originated from the KLT relations \cite{Kawai:1985xq} for tree-level string amplitudes. The celebrated CHY formalism  \cite{Cachazo:2013gna,Cachazo:2013hca,Cachazo:2013iea} extended the tree-level double-copy relation to other theories. Later, ambitwistor string theories \cite{Mason:2013sva,Geyer:2014fka,Geyer:2015jch,Geyer:2019hnn} reproduced the CHY formalism and further clarified the loop-level double copy. In addition, loop-level scattering integrands can be explored through the forward limit of tree-level scattering integrands \cite{He:2015yua,Cachazo:2015aol,Cao:2025ygu}. One can also study the field-theory double copy by taking the low-energy limit of conventional superstring or bosonic-string amplitudes \cite{He:2017spx,Mafra:2017ioj,Mizera:2019blq,Casali:2020knc,Britto:2021prf,Balli:2024wje}.
Although recent advances in quantum field theory, such as five-loop calculations \cite{Bern:2018jmv} and the classical-solution double copy \cite{Monteiro:2014cda}, have extended beyond the current reach of perturbative string theory, string amplitudes continue to serve as an important laboratory for studying scattering amplitudes.

It is worth noting that Eq.~\eqref{eq:double-copy} does not make the internal loop states manifest. In contrast, the BCJ numerators $N_i$ and $\tilde{N}_i$ are functions of external states. We therefore argue that research on the perturbative double copy has traditionally focused on correspondences between external on-shell states. However, studies of the low-energy limit of one-loop type II string amplitudes in Ref.~\cite{Tourkine:2012vx}, together with related work on ambitwistor strings \cite{Geyer:2015jch},
have opened new directions to explore the double-copy structure of internal states exchanged at one loop.\footnote{This decomposition has also been extended to two-loop amplitudes of ambitwistor strings \cite{Geyer:2019hnn}. } Extending these ideas to bosonic and heterotic string theories is the central theme of this paper.
Furthermore, by adopting the chiral-splitting formalism \cite{DHoker:1988pdl,DHoker:1989cxq}, one-loop closed-string integrands exhibit a manifest double-copy relation between the external and internal states of the left- and right-moving sectors, which we study extensively in this paper.      	
	
On the other hand, the running of coupling constants is among the most important predictions of 
quantum field theory (QFT).  
A standard one-loop QCD beta-function formula is given by \cite{Gross:1973id,Politzer:1973fx}:
\begin{equation}
\label{eq:beta-field}
\beta=-\frac{\,g^2\,}{16\pi^2}
\(\!\frac{11}{3}C_{v}-\frac{2}{3}n_fC_{f}-\frac{1}{6}n_sC_{s}\!\)\!,
\end{equation}
where $g$ is the gauge coupling constant, while $n_X$ and $C_{X}$ with
$X=v,f,s$ denote the numbers of vectors (spin-$1$), fermions (spin-$\frac{1}{2}$), and scalars (spin-$0$) and the corresponding quadratic Casimir operators of their representations, respectively.\
The non-Abelian nature of QCD gives a negative beta function, leading to the celebrated asymptotic freedom of the running gauge coupling. This provides an ideal test of whether the decomposition of internal loop states yields the correct field-theory limits.
Several other works in the literature   \cite{Minahan:1987ha,DiVecchia:1996uq,DiVecchia:1996iz} have also derived the gauge beta function from the low-energy limit of string amplitudes. However, these calculations are restricted to specialized string models. Due to the difficulty and complexity of spin-structure summation, previous works usually relied on unbroken supersymmetry. This problem was alleviated by Refs.~\cite{Berg:2016wux,Berg:2016fui}, where spin-structure summation under orbifold compactification with different nonzero amounts of unbroken supersymmetry was studied\footnote{Their main idea is reviewed at the end of Appendix \ref{sec:spin-sum}.}.  
In contrast, internal-loop-state decomposition offers a significant advantage: it directly yields a universal result valid for arbitrary string models.

Moreover, given the non-Abelian nature of general relativity (GR), it is natural to investigate loop corrections to the gravitational beta function arising from the graviton (spin-2), dilaton (spin-0 scalar), antisymmetric tensor field (spin-0 axial scalar in 4D), and gravitino (spin-3/2). However, the complicated Feynman rules make such loop-level calculations difficult in quantum field theory. The loop-exchanged states are not transparent in the traditional loop-level double copy shown in Eq.~\eqref{eq:double-copy}. Fortunately, as we shall see, these internal loop states appear as double copies once the chiral-splitting formalism and internal-loop-state decomposition are used. 

Similarly, one may expect gravity to induce corrections to the gauge beta function proportional to the gravitational constant $\kappa$, where $\kappa^2 = 32\pi G$ and $G$ is the Newtonian gravitational constant. Since the gauge-coupling beta function in four dimensions is dimensionless, one would naively expect the following general form:
\begin{align}
\label{eq:beta-function-general}
\beta_{\text{gauge}} =-\frac{g^2}{\,16\pi^2\,}\!\(\!\frac{11}{3}C_{v}^{}\!-\!\frac{2}{3}n_f^{}C_{f}^{}
\!-\!\frac{1}{6}n_s^{}C_{s}^{}\!\)\!+\!\frac{\,\kappa^2 \mu^2\,}{16\pi^2} a\,,	
\end{align}
where $a$ is a constant to be determined by one-loop calculations. A nonzero  $a$ would imply power-law running of the gauge coupling due to quantum gravitational effects. The possibility of such gravity-induced power-law corrections in Eq.~\eqref{eq:beta-function-general} was first proposed by Robinson and Wilczek \cite{Robinson:2005fj}, leading to a long-standing debate  \cite{Pietrykowski:2006xy,Toms:2007sk,Ebert:2007gf,He:2010mt,Anber:2010uj,Ellis:2010rw}. We address this issue in the present work using internal-loop-state decomposition and the double copy.

This paper is organized as follows. In Section \ref{sec:ILDC}, 
the one-loop scattering integrand is systematically decomposed in terms of internal loop states. The bosonic string is studied in Section \ref{sec:bstring}, and the heterotic string is studied in Section \ref{sec:H-string}. We do not repeat the derivations in Refs.~\cite{Tourkine:2012vx,Geyer:2015jch}, although their main results are included in the decomposition of the supersymmetric right movers of the heterotic string.
For the bosonic left movers of the heterotic string, the fermionic realization is discussed separately. Applications to beta-function calculations are explored in Section \ref{sec:app-beta}. We present the general structure of three-gluon and three-graviton one-loop amplitudes in Section \ref{sec:3pt-1loop} and show how to extract Feynman diagrams from different regions of moduli space in Section \ref{sec:string-d-moduli}. We then compute explicit examples, including the beta function of the gauge coupling in Section \ref{sec:gauge-beta}, the gravitational beta function in Section \ref{sec:gravi-beta}, and possible corrections to the gauge coupling in Section \ref{sec:gravi-gauge}. Finally, the conclusions are summarized in Section \ref{sec:summary}.

\section{Decomposition of the one-loop scattering integrand}
\label{sec:ILDC}
\subsection{Bosonic string}
\label{sec:bstring}
\subsubsection{Structure of one-loop amplitude}
\label{sec:bstring-one-loop}
We first generalize the decomposition method of Refs.~\cite{Tourkine:2012vx,Geyer:2015jch} to bosonic closed-string amplitudes. 
The unintegrated and integrated vertex operators of the graviton are given by
\begin{align}
	V_i & =\zeta\cdot \partial X(z) \xi\cdot \bp X(\bz) e^{ik\cdot X(z_i,\bz_i)}, \\
	U_i & =\int d^2 z \zeta\cdot \partial X(z) \xi\cdot\bp X(\bz)e^{ik\cdot X(z_i,\bz_i)},
\end{align}
where $\{z,\bz\}$ are the complex coordinates on the world-sheet, $X^\mu$ are the spacetime coordinates, and $\zeta$ and $\xi$ are polarization vectors\footnote{Here we use different symbols to distinguish left and right movers.}.

Thus the one-loop scattering amplitude is given by
\begin{equation}
	\mathcal{M}_{(1)}^n=\int_{\mathcal F} \frac{d^2\tau}{\tau_2^{d/2}| \eta(\tau)|^{48}} Z_e(\tau,\bar \tau) \langle V_1 U_2 ...U_n \rangle,
\end{equation} 
where we fix $z_1=0$ without loss of generality. Here $\mathcal F$ denotes the fundamental region of the $\mathbb{SL}(2,Z)$ modular group, and $\tau=\tau_1 + i \tau_2$ is the modulus of the torus world-sheet. The factor $\frac{Z_e(\tau,\bar \tau)}{\tau_2^{d/2} |\eta(\tau)|^{48}}$ is the total partition function on the torus, and $Z_e$ counts the additional contributions from Kaluza-Klein and winding modes due to compactification. For instance, consider a string propagating in $R^{1,d-2}\otimes T^{26-d}$ spacetime with radii $R_i,i=1,...,(26-d)$ and 
\begin{equation}
	Z_{e}(\tau,\bar \tau) =\prod_{i=1}^{26-d} \sum_{n,w} e^{-\pi \tau_2 (\frac{\alpha' n_i^2}{R_i^2}+\frac{w_i^2 R_i^2}{\alpha'})+2\pi i n w\tau_1}.
\end{equation}

However, the correlation functions $\langle V_1 U_2 ...U_n \rangle$ cannot in general be written as products of holomorphic and anti-holomorphic functions. This is because the scalar propagator on the torus includes a non-holomorphic part,
\begin{equation}
	G(z,\bz) = G(z) + \bar G(\bz) + \frac{2\pi y}{\tau_2},
\end{equation}
where $z=x+iy,x,y\in R$, and we define the chiral scalar propagator as $G(z)=-\log \theta_1(z,\tau)$. The non-holomorphic part $\frac{2\pi y}{\tau_2}$ obscures the factorization structure at one loop. To restore this factorization, we adopt the chiral-splitting formalism \cite{DHoker:1989cxq,DHoker:1988pdl} and write 
\begin{equation}
	\langle V_1 U_2 ...U_n \rangle= \tau_2^{d/2}\int d^{d}l f(\zeta_i,k_i,l,z,\tau) \bar f(\xi_i,k_i,l,\bz,\bar \tau) |\mathcal J_n|^2.	
\end{equation}
Here we introduce the loop momentum $l$. The chiral Koba-Nielsen factor $\mathcal J_n$ is given by
\begin{align}
	\mathcal{J}_n=&\exp[\frac{\pi i \tau \alpha' l^2}{2} +\pi i \sum_i l\cdot k_i z_i +\sum_{i<j}\frac{\alpha'k_i \cdot k_j}{2}  \log \theta_1(z_{ij},\tau)],
\end{align}
and $f(z)$ ($\bar f(\bz)$) which is a function of $\partial^{1,2} G(z)$ ($\partial^{1,2}\bar G(\bz)$), the momenta $k_i$, and the polarization vectors $\zeta_i$ ($\xi_i$) of the external states. In addition, to ensure that
\begin{equation}	
\int dl^d \mathcal J_n \bar{\mathcal J}_n = \int dl^d e^{-\pi\tau_2 \alpha'l^2+\dots} 	
\end{equation}
is convergent, the loop momentum $l$ is always taken to be Euclidean. A systematic comparison between the loop momentum introduced by the chiral-splitting formalism and its field-theory counterpart can be found in Ref.~\cite{Tourkine:2019ukp}.  Thus the double-copy structure is restored after introducing the loop momentum $l$, at the cost of making modular symmetry non-manifest. The loop-level KLT relation \cite{Stieberger:2023nol} was also derived within the chiral-splitting formalism.

Next, we turn to the world-line limit $\tau_2 \rightarrow \infty$. Since we are interested in the model-independent part, the factor $Z_e$ is neglected in the following calculation. It is conventional to expand the above integrand in terms of $q=e^{i\pi  \tau}$ and $\bar q=e^{-i \pi  \tau}$, whose coefficients of $q^n\bar q^m$ count the degrees of freedom of the internal loop states, as shown in Table.~\ref{table:qbq-expansion}. 

There is a subtle issue in identifying which internal loop state, for a given loop in a Feynman diagram or hole on the world-sheet, corresponds to the $q\bar q$ expansion. Recall that, in the operator formalism \cite{Green:2012pqa}, the one-loop open-string planar amplitude is proportional to 
\begin{equation}
	A_n^1 \sim \int dp^d Tr[\Delta_1 V_1(1) \Delta_2 V_2(1) \Delta_3 ... V_n(1) ],
\end{equation}
where $\Delta_i=\int_0^1 x_i^{L_0-2} dx_i$ is the world-sheet propagator. However, using $x^{L_0}V(1)=V(x)x^{L_0}$, the above equation becomes
\begin{equation}
	A_n^1 \sim \int dp^d \prod_i \int_0^1 dx_i Tr[ V_1(x_1)  V_2(x_1x_2) ... V_n(x_1 x_2...x_m) w^{L_0-2} ], w=x_1 x_2...x_m.
\end{equation}
In this way, a special propagator is chosen, even though all propagators in the loop play the same physical role. Moreover, the partition function arises from the chosen propagator after expanding the world-sheet fields into different modes and performing the loop integral $\int dp^d$. Consequently, the propagator corresponding to the partition function is the one near the vertex operator fixed at $z=0$ or $\bar z=0$, which we simply refer to as an internal loop state in this paper.  

Although the physical meaning is clearer in the operator formalism, the path-integral formalism is better suited for calculations because of conformal symmetry. We will not delve into the details of this traditional method.
In addition, the chiral-splitting path-integral formalism allows us to treat the holomorphic and anti-holomorphic parts separately.

\begin{table}
	\centering
	\begin{tabular}{|c|c|}
		\hline
		$q^{-2} \bar q^{-2}$ & tachyon \\
		\hline
		$q^{0} \bar q^{0}$ & massless states \\
		\hline
		$q^{2n} \bar q^{2m},m=n>1$ & Regge states \\
		\hline
	\end{tabular}
	\caption{Unphysical states with $n\neq m$ are removed by the level matching condition, which is realized by the real part integral of $\tau_1$.}
	\label{table:qbq-expansion}
\end{table}	

For later convenience, we define the chiral integrand as 
\begin{equation}
	W(z)=f(\zeta_i,k_i,l,\tau) \mathcal J_n=\sum_m W_m(z) q^m,
\end{equation}
which is analogous to the planar-diagram integrand of the open-string scattering amplitude. In addition, we emphasize that in the original formulation of Refs.~\cite{Tourkine:2012vx,Geyer:2015jch}, the factors $\mathcal{J}_n$ and $\bar{\mathcal{J}}_n$ are absent. We will see, however, that $\mathcal{J}_n$ and $\bar{\mathcal{J}}_n$ must be included for the bosonic string and for the 26-dimensional left movers of the heterotic string.

\subsubsection{Decomposition of the scattering integrand}
\label{sec:bstring-d-integrand} 
In this section, we focus on the massless states, so $Z_e$ is neglected.
Recall that the Dedekind eta function has the following expansion
\begin{equation}
	\frac{1}{\eta(\tau)^{24}}=\frac{1}{q^2} + 24+ \mathcal{O}(q^2),
\end{equation}
Since the partition function counts the physical spectrum, we can split the integrand in terms of states exchanged in the loop,
\begin{align}
	I_{L}^{\text{tachyon}}&=W_0(z), \\
	I_{L}^{\text{massless}}&=W_2(z)+24 W_0(z).
\end{align}
Since the tachyon is a scalar, one might naively conjecture that  
\begin{align}
	I_L^{\text{scalar}}&=n_s W_0(z), \\
	I_{L}^{\text{vector}}&=W_2(z) + (d-2)W_0(z).
\end{align}
Here we replace the number of scalars, $26-d$, by a general number $n_s$ because we are interested only in the field-theory limit.
However, it turns out that the above conjecture is valid for pure massless graviton scattering amplitudes.

It is worth noting that any graviton vertex operator can be generated from the multilinear part of 
\begin{equation}
	\tilde V= e^{ik.X+i\zeta_1.\partial X + i\xi_1.\bp X  }=e^{i\tilde{k} \cdot X},
\end{equation}
where we define
\begin{equation}
	\tilde{k}=k+\zeta_1\partial+\xi_1\bp .
\end{equation}
Thus 
\begin{align}
	\tilde W(z) &= \exp[\frac{\pi i \tau \alpha' l^2}{2} +\pi i \sum_i l\cdot k_i z_i +\sum_{i<j}\frac{\alpha'\tilde k_i \cdot \tilde k_j}{2}  \log \theta_1(z_{ij},\tau)], \nonumber \\
	&=\mathcal J_3 \exp[\sum_{i<j}\frac{\alpha'\tilde k_i \cdot \tilde k_j}{2}  \log \theta_1(z_{ij},\tau)].
\end{align} 
Since $X^\mu$ are free fields on the world-sheet, we can rewrite $\tilde W(z)$ as 
\begin{equation}
	\exp[\sum_{i<j}\frac{\alpha'\tilde k_i \cdot \tilde k_j}{2}  \log \theta_1(z_{ij},\tau)]=\prod_\mu T_\mu,	
\end{equation}
where 
\begin{equation}
	T_\mu=\exp[\sum_{i<j}\frac{\alpha'\tilde k_i^\mu\tilde k_j^\mu}{2}  \log \theta_1(z_{ij},\tau)]=\sum_n T_\mu^{(n)} q^n.
\end{equation}
Here $T_\mu$ can be identified as the contribution of the nonzero modes of $X^\mu$. It is straightforward to see that $X^{\mu=1,..,d}$ is associated with vector states, while $X^{d+1,...,26}$ corresponds to scalar states. In this way,
\begin{align}
	I_L^{\text{vector}}&=\sum_{\mu=1}^d (T^{(2)}_\mu)|_{\text{multi-linear}} + (d-2)W_0, \\
	I_L^{\text{scalar}}&=\sum_{\mu=d}^{n_s+d} (T^{(2)}_\mu)|_{\text{multi-linear}} + n_s W_0.
\end{align} 
It is straightforward to see that the above formula reduces to our naive conjecture if and only if $ \sum_{\mu=d}^{n_s+d}  (T^{(2)}_\mu)|_{\text{multi-linear}}=0$.  A similar decomposition can be applied to the anti-holomorphic right movers. 

\subsection{Heterotic string}
\label{sec:H-string}
\subsubsection{Scattering amplitude with even spin structure}
\label{sec:Hstring-oneloop}
In this subsection, we study the scattering amplitude of the heterotic string in $R^{1,3}\otimes T^6$. In our convention, the right movers are chosen to be the RNS superstring \cite{Polchinski:1998rr}, and the left movers are chosen to be the bosonic string. The results of Refs.~\cite{Tourkine:2012vx,Geyer:2015jch} will be revisited when the right movers are discussed. 

The integrated vertex operators of the gluon with ghost numbers $-1$ and $0$ are given by
\begin{align}
	U^{0,1}_i&=\int d^2 z j_{a_i}(z) (i\xi_i\cdot \bar \psi )e^{ik_i.X}, \\
	U^{1,1}_i&=\int d^2 z j_{a_i}(z)(i\xi_i\cdot\partial X+k_i\cdot\bar \psi \xi_i\cdot\bar \psi)e^{ik.X},
\end{align}
where $\psi^\mu$ are the world-sheet fermion fields and $j_{a_i}(\bz)$ are currents of the WZW model \cite{DiFrancesco:1997nk}. In the fermionic realization, the currents are given by
\begin{equation}
	j_{a_i}=f_{a_i b c} \psi_b  \psi_c
\end{equation}
in the fermionic realization. Similarly, the graviton vertex operators are given by
\begin{align}
	U^{0,2}_i&=\int d^2 z (i\zeta_i\cdot\partial X) (i\xi_i\cdot\bar \psi )e^{ik_i\cdot X}, \\
	U^{1,2}_i&=\int d^2 z (i\zeta_i\cdot\partial X)(i\xi_i\cdot\bp X+k_i\cdot\bar \psi \xi_i\cdot\bar \psi)e^{ik\cdot X}.
\end{align}
For unintegrated vertex operators, we continue to use the symbol $V$.

Any $n$-point one-loop scattering amplitude with even spin structures in the chiral-splitting formalism \cite{DHoker:1988pdl,DHoker:1989cxq} is given by 

\begin{align}
	\label{eq:3p1L-het}
	\mathcal{A}_{n}^{(1)} = &\, \frac{~g^n_{10}\! }{\,(2\pi)^3V\,}\!\!
	\!\sum_{\alpha_L,\beta_L}\!\sum_{\alpha_R,\beta_R}\!\!
	\int_{\!\mathcal F}^{} \!\!
	\frac{\,d^2\tau  \prod_i d^2 z_i^{}\,}{(4\pi^2 )^2 V_{\text{CKV}}} \int d^4l
	P_L(\alpha_L,\beta_L,\tau)P_R(\alpha_R,\beta_R,\bar\tau) 
	\nonumber \\
	& \times\! 
	Z_{T^6}(\tau,\bar \tau)
	W(\alpha_L,\beta_L,\tau,l) \bar W(\alpha_R,\beta_R,\bar z,\bar \tau,l)\,,
\end{align}
where $P_L^{}$ and $P_R^{}$ are the left- and right-moving partition functions  
for the given spin structures $(\alpha_L,\beta_L)$ and $(\alpha_R,\beta_R)$, respectively (excluding compactification), and $Z_{T^6}$ is the extra factor induced by $T^6$ compactification.  
$\mathcal F$ denotes the fundamental region under modular transformations and is usually chosen as 
$|\tau|\!\!\geqq \! 1$ and $-\frac{1}{2} \!<\! \tau_1^{} \!\leqq\! \frac{1}{2}$,
where the moduli parameter $\tau=\tau_1^{}+ i\tau_2^{}$.\ 
The quantity $V=(2\pi R)^6$ denotes the volume of the extra-dimensional space.\ 
The quantities $W$ and $\bar W$ are the left- and right-moving effective correlation functions
in the chiral-splitting formalism, with
\begin{equation}
	W(\alpha_L,\beta_L,\tau,l) \bar W(\alpha_R,\beta_R,\bar z,\bar \tau,l)=\langle V_1^{1,\star} U_2^{1,\star} ...U_n^{1,\star}\rangle_{[\alpha,\beta]}^{} ,\star=1,2.
\end{equation}
To decompose the internal loop states, we expand $W^L$ and $W^R$ in terms of $q$ and $\bar{q}$,
\begin{align}
	W(\alpha_L,\beta_L,\tau,l)&=\sum_{n}W_n(\alpha_L,\beta_L,\tau,l) q^n, \\
	\bar W(\alpha_R,\beta_R,\tau,l)&=\sum_{n}\bar W_n(\alpha_R,\beta_R,\bar \tau,l) \bar q^n.	
\end{align}

We now briefly review the approach of Refs.~\cite{Tourkine:2012vx,Geyer:2015jch}. Recall that the partition functions of the right movers can be expanded as 
\begin{align}
	P_R^{}(0,0)&=\frac{1}{\bar q}+8+\mathcal{O}(\bar q), \\
	P_R^{}(0,1/2)&=-\frac{1}{\bar q}+8+\mathcal{O}(\bar q), \\
	P_R^{}(1/2,0)&=-16+\mathcal{O}(\bar q).
\end{align}
The fermions lie in the R sector $(1/2,0)$, while the bosons lie in the NS sector $(0,0 \text{ or } 1/2)$. The term $\frac{1}{\bar q}P_R^{}(0,0 \text{ or }1/2)|_{\bar q^0}$ should correspond to the exchange of the unphysical scalar tachyon in the loop. It was therefore argued in Refs.~\cite{Tourkine:2012vx,Geyer:2015jch} that 
\begin{equation}
	I_R^{\text{scalar}}=\frac{n_s}{2} (\bar W_0(0,0)+\bar W_0(0,1/2)=n_s \bar W_0(0,0).
\end{equation}
where $n_s$ is the number of real scalar particles.
In the second equation, we used the fact that the tachyon scalar must be canceled, with $\bar W_0(0,0)=\bar W_0(0,1/2)$.
Although this result is correct, a more transparent and persuasive proof can be obtained by comparing the $\mathbb Z_3$ orbifold untwisted partition functions \footnote{The generator of $\mathbb Z_3$ is chosen to be 
	\begin{equation}
		{\tt g}   = e^{i2\pi (\alpha_1^{} J_{45}^{}+\alpha_2^{} J_{67}^{}+\alpha_3^{} J_{89}^{})} 
		e^{i 2\pi (\beta_1^{} H_{12}^{}+\beta_2^{}H_{34}^{}+\beta_3^{}H_{56}^{})}.
	\end{equation}
	where $\{J_{ij}^{}\}$ denote the rotation generators in spacetime and 
	$\{H_{ij}^{}\}$ are the generators of the Cartan subalgebra of 
	$SO(32)$
	which act as rotation operators in the internal space.
} 
\begin{align}
	\frac{1}{3} \sum_h P_R^{\tt e \tt h}(0,0)&=\frac{1}{\bar q}+2+\mathcal{O}(\bar q), \\
	\frac{1}{3} \sum_h P_R^{\tt e \tt h}(0,1/2)&=-\frac{1}{\bar q}+2+\mathcal{O}(\bar q), \\
	\frac{1}{3} \sum_h P_R^{\tt e \tt h}(1/2,0)&=-4+\mathcal{O}(\bar q).
\end{align}
By counting the degrees of freedom, we further conclude that 
\begin{align}
	I_{R}^{\text{vector}}=&\frac{1}{2}(\bar W_1(0,0)-\bar W_1(0,1/2) \nonumber \\
	& \frac{d-2}{2}(\bar W_0(0,0)+\bar W_0(0,1/2) \nonumber \\
	=&\bar W_1(0,0)+(d-2)\bar W_0(0,0), \\
	I_R^{\text{fermion}}=& -2 n_f \bar W_0(1/2,0).
\end{align}
where $d$ is the dimension of flat spacetime and $n_f$ is the number of spacetime fermions in four dimensions. In the toy models studied in this paper, we always have $n_v=1$.

It is noteworthy that, aside from $\bar W_1(0,0)$, all the terms involving $\bar W_0(0 \text{ or } 1/2,0)$ in $I_R^{\star},\star \in \{ \text{vector, scalar and fermion} \}$ are proportional to the degrees of freedom of the corresponding states. This observation can be understood as follows. The terms $\pm \frac{1}{\bar q}$ indicate the vacuum in the NS sector. The factor $\bar W_1(0,0)$ counts the exchange of massless states created by the inserted vertex operators acting on the NS vacuum, while $\bar W_0(0 \text{ or } 1/2,0)$ counts the exchange of intrinsic massless modes propagating on the world-sheet. 
In addition, the Koba-Nielsen factor $\bar{\mathcal J}(\bz)$ contributes only to $\bar W_n(\alpha_R,\beta_R)$ for $n\geq 2$. Thus, our formulas yield exactly the same result as Refs.~\cite{Tourkine:2012vx,Geyer:2015jch} where $\bar{\mathcal J}(\bz)$ is not considered.

In the bosonic realization, the left-moving sector can be identified with the bosonic string. The decomposition of the internal loop states is exactly the same as in the bosonic string, as discussed in Section \ref{sec:bstring}. For the gluon and gluino, it is more convenient to use the fermionic realization, where the partition function of the $SO(32)$ heterotic string is given by
\begin{align}
	P_L^{} (0,0)&= \frac{1}{q^2}+\frac{32}{q}+504+ \mathcal{O}(q), \\
	P_L^{} (0,1/2)&= \frac{1}{q^2}-\frac{32}{q}+ 504+ \mathcal{O}(q), \\
	P_L^{} (1/2,0)&= 65536q^2+ \mathcal{O}(q^3).
\end{align}
Since there is no massless state in the $(1/2,0)$ sector, we focus on the $(0,0)$ and $(0,1/2)$ sectors.
Thus we can identify
\begin{align}
	c e^{ik\cdot X_L}  \sim & \frac{1}{2q^2}( W_0(0,0) + W_0(0,1/2) ) \\
	c  \psi^a e^{ik\cdot X_L}  \sim & 
	\frac{32}{2q}(W_0(0,0)-W_0(0,1/2)) \nonumber \\ 
	&+ \frac{1}{2q}(W_0(0,0)+W_0(0,1/2))  \\
	c\{\partial X^\mu, \psi^a \psi^b \}e^{ik\cdot X_L}  \sim & 
	\frac{32}{2}(W_1(0,0)-W_1(0,1/2)) \nonumber \\
	&+ \frac{1}{2}(W_2(0,0)+W_2(0,1/2)) \nonumber \\
	&+\frac{504}{2}(W_0(0,0) + W_0(0,1/2) )
\end{align}
The scalar tachyon $c e^{ik\cdot X_L}$ is present in the left-moving spectrum but is removed by the level-matching condition in the heterotic string amplitude. The $SO(32)$ tachyon $c \psi^a e^{ik\cdot X_L}$ should be canceled by the GSO projection; hence, we must have
\begin{align}
	W_0(0,0) & =W_0(0,1/2), \\
	W_1(0,0)& =-W_1(0,1/2).
\end{align}
The main difficulty in extracting the gravitational correction is how to separate the contributions of $c \partial X^\mu e^{ik\cdot X_L}$ and $c \psi^a \psi^b e^{ik\cdot X_L}$. 

On the other hand, the fermion propagator $S_{\delta}(z,\tau)$ with even spin structures is given by
\begin{equation}\label{eq:S-exact}
	S_{\delta}(z,\tau)=\frac{\theta_\delta(z,\tau)\theta'_1(0,\tau)}{\theta_\delta(0,\tau)\theta_1(z,\tau)},\delta=2,3,4.
\end{equation}
If $\tau_2\rightarrow + \infty$ and $y\sim \log(\tau_2)$ or $y\simeq \tau_2$, 
\begin{align}
	S_2(z,\tau)&\simeq  \pi \cot(\pi z) - 4\pi \sin(2\pi z) q^2+\mathcal{O}(q^3),  \\
	S_3(z,\tau)&\simeq  \pi \csc(\pi z) -4\pi \sin(\pi z)(q-q^2) +\mathcal{O}(q^3), \\
	S_4(z,\tau)&\simeq  \pi \csc(\pi z) +4\pi \sin(\pi z)(q+q^2) +\mathcal{O}(q^3).
\end{align}
Moreover, if $y\sim \frac{1}{\tau_2}$, then for any spin structure $\delta$,
\begin{equation}
	S_\delta(z,\tau)\sim \frac{1}{z}.
\end{equation}

One can see that the quadratic Casimir operator arises from the limit $y_{ij}\rightarrow \infty,i\neq j$:
\begin{equation}
	4\pi^3i C_{SO(32)}=32\times 8\pi^3 i - 1\times 16\pi^3 i +  504\times 0.
\end{equation}
Here we discard terms proportional to $\frac{1}{w_{21}}$ or $\frac{1}{w_{32}}$, where $w_{ij}=e^{2\pi i z_{ij}}$, because they can be removed by the real-part integral $\int dx_2 dx_3$ after being combined with the left movers.

With the above evidence, we find that the conventional tropical limit \cite{Tourkine:2013rda}, where $q\rightarrow0$ and $w\rightarrow 0$, provides only the color part of the left movers. The $q^0$ term $504=8+496$ counts the massless states in the left-moving spectrum: $8$ counts $c\partial X^\mu e^{i k \cdot X_L}$, and $496=\frac{32\times 31}{2}$ counts $c\psi^a  \psi^b e^{ik\cdot X_L}$. We therefore argue that
\begin{align}
	I_L^{\text{scalar}} 
	=&n'_s W_0 + \sum_{\mu=d}^{n'_s+d} (T^{(2)}_\mu), \\
	I_L^{\text{color}} =& W_2(S) + n_c W_1 +\frac{n_c(n_c-1)}{2}W_0,   \\
	I_L^{\text{vector}} =& (d-2)W_0 + \sum_{\mu=1}^d (T^{(2)}_\mu).
\end{align}
Here we define $n_c=32$ for $SO(32)$ heterotic string theory and split $W_2$ into 
\begin{equation}
	W_2=W_2(S) + W_2(G),
\end{equation}
where $W_2(S)$ and $W_2(G)$ denote the coefficients of $q^2$ from the fermion propagator $S_{\alpha_L,\beta_L}(z,\tau)$ and the chiral boson propagator $G(z)$, respectively. 

As in the bosonic string,
\begin{equation}
	W_2(G)=\sum_{\mu=1}^d (T^{(2)}_\mu)|_{\text{multi-linear}} + \sum_{\mu=d}^{n'_s+d} (T^{(2)}_\mu)|_{\text{multi-linear}}.
\end{equation}
Hence,
$W_2(G)$
is universal and has a unique physical meaning. It is also interesting that $W_2(G)$ is absent from the supersymmetric right movers of heterotic string theory and from other superstring theories. We therefore argue that, on the one hand, it denotes a correction from the $\alpha' F^3$ effective interaction in the open string planar one-loop amplitude. On the other hand, it describes the quantum correction from the gravitational sector to loop-scattering processes in heterotic string theory. In particular, when all external states are gluons or gluinos, this term shows how gravity affects non-gravitational perturbative processes. Such a loop correction is not present as a simple torus diagram in type II A/B or type I superstring theory\footnote{Branes and open-closed diagrams must be considered.}. This is the reason why we are able to clarify the gravitational correction to the gauge beta function in Section \ref{sec:gravi-gauge}. Moreover, $W_2(G)$ is also missing in the ambitwistor string amplitude \cite{Geyer:2015jch}.

It is also worth noting that if all external states are gluons or gluinos, $W_0$ is always zero for the $SO(32)$ heterotic string. This observation guarantees that the right-moving integrand $W^R$ provides the correct quadratic Casimir operator for the gauge sector.   

As a consistency test, we can consider the $E8\times E8$ heterotic string and focus on the first $E8$ gauge group,
\begin{align}
	P_L(0,0)&=\frac{1}{q^2}+\frac{16}{q}+376+\mathcal{O}(q), \\
	P_L(0,1/2)&=\frac{1}{q^2}-\frac{16}{q}+376+\mathcal{O}(q), \\
	P_L(1/2,0)&=256 +\mathcal{O}(q). 
\end{align}
It is straightforward to check that 
\begin{equation}
	4\pi^3i \times C_{E8} = 8i\pi^3\times 16-16 i\pi^3\times 1+376\times 0 -256 i\pi^3 /2 .
\end{equation}
Since there are no $q^{-2}$ or $q^{-1}$ terms in $P_L(1/2,0)$, $I_L^{\text{scalar}}$ and $I_L^{\text{vector}}$ are the same as in the $SO(32)$ heterotic string.

\subsubsection{Scattering amplitude with odd spin structure}
\label{sec:Hstring-odd}
For the odd spin structure, there is a supermodulus (the zero mode of the two-dimensional gravitino on the world-sheet) and a superconformal Killing vector on the torus. Therefore, we must insert a picture-changing operator and replace one of the vertex operators by its partner with ghost number $0$,

\begin{align}
	\mathcal{A}_{n}^{(1)}(\text{odd}) = &\, \frac{~g^n_{10}\! }{\,(2\pi)^3V\,}\!\!
	\!\!\sum_{\alpha_R,\beta_R}\!\!
	\int_{\!\mathcal F}^{} \!\!
	\frac{\,d^2\tau  \prod_i d^2 z_i^{}\,}{(4\pi^2 )^2 V_{\text{CKV}}} \int d^4l
	P_L(1/2,1/2,\tau)P_R(\alpha_R,\beta_R,\bar\tau) 
	\nonumber \\
	& \times\! 
	Z_{T^6}(\tau,\bar \tau)
	\langle \mathcal{X} V_1^{0,\star} U_2^{1,\star} ...U_n^{1,\star}\rangle_{[\alpha,\beta]}^{} ,\star=1,2\,,
\end{align} 
where $\mathcal{X}$ is the picture-changing operator that is given by
\begin{equation}
	\mathcal{X}=\delta(\beta) S_F.
\end{equation}
The details of these definitions can be found in the standard textbook \cite{Polchinski:1998rr}.

The fermionic propagator on the torus with odd spin structure is given by
\begin{equation}
	S_{1}(z,\tau)=-\partial G(z,\bz,\tau)=-\partial \log\theta_1(z,\tau) + \frac{2\pi i}{\tau_2} y
\end{equation}
and $\frac{2\pi i}{\tau_2} y$ reflects the existence of zero modes of $\psi^\mu$ on the torus with odd spin structure. 
It is known that such CP-violating amplitudes reproduce the Chern-Simons term in SYM and SUGRA. 
Notice that
\begin{equation}
	P_L(1/2,1/2,\tau)=1,
\end{equation}
which reflects the fact that only massless fermions are exchanged in these diagrams, as in quantum field theory. The internal-loop decomposition is trivial in the odd spin structure.

We end this section by summarizing the double-copy relation for selected internal states of the heterotic string in Table.~\ref{table:double-copy}. It is straightforward to extend this table of double copies of internal loop particles to other string theories.

\begin{table}
	\centering
	\begin{tabular}{|c|c|c|c|}
		\hline
		& $I_L^{\text{vector}}$  &  $I_L^{\text{scalar}}$ & $I_L^{\text{color}}$  \\
		\hline
		$I_R^{\text{vector}}$& ${h,B,\Phi}$ & $A'$ & $A^a$  \\
		\hline
		$I_R^{\text{scalar}}$& $A''$ & $\phi$  & $\phi^a$  \\
		\hline
		$I_R^{\text{fermion}}$& $\Psi^\mu$ & $\psi$  & $\lambda^a$  \\
		\hline
	\end{tabular}
	\caption{The double copies of internal loop particles are analogous to the conventional double copies of external particles.}
	\label{table:double-copy}
\end{table} 

\section{\hspace*{-2.5mm} Applications to beta functions}
\label{sec:app-beta}
\subsection{\hspace*{-2.5mm}Three-point one-loop amplitudes of the heterotic string}
\label{sec:3pt-1loop}
In this section, we consider the heterotic string with naive $T^6$ compactification with radii $R_i,i=4,...,9$. First, the tree-level three-point scattering amplitude can be written as:
%
\begin{align}
	\mathcal A_{3(0)}^{} &=  \ii\hs \gh\hs f^{a_1^{}a_2^{}a_3^{}}\hs 
	\xi_1^\mu \xi_2^\nu \xi_3^\si V^{}_{\mu \nu \si}  
	\nn\\
	&\equiv \ii\hs \gh \hs f^{a_1^{}a_2^{}a_3^{}} \!
	\left[(\xi_1^{}\hsm\!\cdot\hsm \xi_2^{})(\xi_3^{} \!\cdot\hsm k_{1}^{})\!+\!
	(\xi_2^{}\hsm\!\cdot\hsm \xi_3^{})(\xi_1^{} \!\cdot\hsm k_{2}^{}) \!+\!
	(\xi_3^{}\hsm\!\cdot\hsm \xi_1^{})(\xi_2^{} \!\cdot\hsm k_{3}^{}) \right]
	\hsm, 
	\label{eq:A3(0)}
\end{align}
%
where $f^{abc}$ are the structure constants of the non-Abelian gauge group and 
$V^{\mu \nu \si} \!\!=\!\eta^{\mu \nu}k_{1}^\si \hsm +\hsm \eta^{\nu \si}k_{2}^\mu 
\hsm +\hsm\eta^{\si \mu}k_{3}^\nu\hs$ 
is the kinematic factor for the cubic gauge interactions.\ 
Here, $\gh\hs$ denotes the 10-dimensional Yang-Mills gauge coupling.
The tree-level three-point amplitude of three gravitons can be obtained by the double copy,
\beqs 
\begin{align}
	\mathcal M_{3(0)}^{} &=  \ii\hs \kappa \hs 
	\xi_1^\mu \xi_2^\nu \xi_3^\si V^{}_{\mu \nu \si}
	\hs 
	\zeta_1^\rho \zeta_2^m \zeta_3^n V^{}_{\rho m n}. 
\end{align}
\eeqs

Next, we study the three-point one-loop amplitude.\ Any on-shell three-point scattering amplitude 
at loop level vanishes identically because the on-shell conditions for massless external states require trivial Mandelstam variables 
$\,k_i^{} \!\cdot\hsm k_j^{} \hsm =\hsm 0\hs,i,j=1,2,3$.\ 
To compute the physical beta function from 
the one-loop three-point amplitude, we follow the infrared regularization introduced in Ref. \cite{Minahan:1987ha}\footnote{The same infrared regularization was also employed in Refs.~\cite{Berg:2016wux,Berg:2016fui} in their study of one-loop three-point amplitudes. Their work focuses on the analytic evaluation of these amplitudes, in particular the summation over spin structures. By contrast, the present work aims to reveal the double-copy structure of the internal loop states. Regarding the infrared regularization itself, there is no substantial difference between the treatment adopted here and that used in those papers.
}: 
\begin{equation}\label{eq:IR-Minahan}
	k_1^{} + k_2^{} + k_3^{} = p \hs, 
\end{equation}
with $p^2\!=\!0\hs$.\ 
We may denote $k_4^{}\!\equiv\!-p$, giving momentum conservation
$k_1^{}\!+\!k_2^{}\!+\!k_3^{}\!+\!k_4^{}\!=\!0\hs$.\ 
The transversality conditions 
 $\xi_j^{} \!\cdot\! k_j^{}\!=\zeta_j \cdot k_j=\!0\hs$ 
hold for $j\!=\!\{1,2,3\}$.\ 
Defining  $s_{ij}^{}\!=\!(k_i^{}\!+\!k_j^{})^2$, we obtain the following kinematic identity:
\beq 
\label{eq:Sij-sum=0}
s_{12}^{}+s_{23}^{}+s_{31}^{} = 0 \, .
\eeq 
These constraints preserve conformal invariance of the correlation function.\  
Effectively, this corresponds to a four-point amplitude with one external state in the soft limit. Additionally, we adopt the axial gauge for all the polarization vectors $\xi_i,i=\{1,2,3\}$, which imposes $\xi_i \cdot p=0$. 

\vs

The one-loop three-gluon scattering amplitude is then given by
\begin{align}
	\label{eq:3p1L-het}
	\A_{3(1)}^{} = &\, \frac{~\hs\gh^3\! }{\,(2\pi)^3V\,}\!\!
	\!\sum_{\alpha_L,\beta_L}\!\sum_{\alpha_R,\beta_R}\!\!
	\int_{\!\mathcal F}^{} \!\!
	\frac{\,\di^2\tau\hs\di^2z_2^{}\hs\di^2z_3^{}\,}{(4\pi^2 )^2} \int d^4l|\mathcal{J}_3|^2
	\nonumber \\
	& \times\! P_L(\alpha_L,\beta_L,\tau)P_R(\alpha_R,\beta_R,\bar\tau) Z_{T^6}(\tau,\bar \tau)
	\langle V_1^{1,1} U_2^{1,1} U_3^{1,1}\rangle_{[\alpha,\beta]}^{}\,,
\end{align}
where the correlation function of vertex operators  
$\langle V_1^{1,1} U_2^{1,1} U_3^{1,1}\rangle_{[\alpha,\beta]}^{}\!$
is derived below in Eq.~\eqref{eq:U1U2U3}.\ 
Here $P_L^{}$ and $P_R^{}$ are the left- and right-moving partition functions  
for the given spin structures\footnote{Since there are two periodic directions on a torus, we should specify two boundary conditions 
	of $\psi^M$.\ Thus, we need to sum over the spin structures $\{\alpha_R^{},\beta_R^{}\}$ of 
	the right-moving part and $\{\alpha_L^{},\beta_L^{}\}$ of the left-moving part independently.\ } $(\alpha_L,\beta_L)$ and $(\alpha_R,\beta_R)$, respectively (excluding compactification), and $Z_{T^6}$ is the extra factor induced by $T^6$ compactification.  
$\mathcal F$ denotes the fundamental region under modular transformations and is usually chosen as 
$|\tau|\!\!\geqq\hsm\! 1$ and $-\frac{1}{2} \hsm\!<\!\hsm \tau_1^{} \hsm\!\leqq\hsm\! \frac{1}{2}\hs$,
where the moduli parameter $\tau=\tau_1^{}+\ii\hs\tau_2^{}$.\ 
The quantity $V=(2\pi R)^6$ denotes the volume of the extra-dimensional space.\ 
Finally, $\mathcal{J}_3$ is the chiral Koba--Nielsen (KN) factor, which is given by
\begin{align}
\mathcal{J}_3=&\exp[\frac{\pi i \tau \alpha' l^2}{2} +\pi i \sum_i l\cdot k_i z_i +\sum_{i<j}\frac{\alpha'k_i \cdot k_j}{2}  \log \theta_1(z_{ij},\tau)]
\end{align}
where $l$ is the four-dimensional loop momentum. 

For the even spin structures $[\alpha_\star,\beta_\star]$, we derive the correlation function of the three 
vertex operators in the chiral-splitting formalism as follows:
\begin{align}
	\label{eq:U1U2U3} 
	\langle V_1^{1,1} U_2^{1,1} U_3^{1,1}\rangle_{[\alpha_\star,\beta_\star]}^{} 
	=& \frac{\alpha'^2}{8} \Big[ \Lambda^1(\bar z)+\Lambda^2(\alpha_R,\beta_R,\bar z)+\Lambda^3(\alpha_R,\beta_R,\bar z) \Big] \nn \\
	&\Phi_{\alpha_L^{}\beta_L^{}}^{}\!\!(T^{a_1},T^{a_2},T^{a_3})   \hs,  
\end{align}
where $\Lambda^1$, $\Lambda^2$ and $\Lambda^3$ are given by 
\begin{align}
	\label{eq:lambda-1}
	\hspace*{-9mm}
	\Lambda^1(\bar z) =&  
	 \alpha' (\xi_1\cdot k_2 (\bar \partial G_{12}-\bar \partial G_{13})-2 \pi i  l\cdot\xi_1) \nn \\
	 &(\xi_2\cdot k_3 (\bar \partial G_{12}+\bar \partial G_{23})-2 \pi i  l\cdot \xi_2) \nn \\
	 &\left(\xi_3\cdot k_1 \left(\bar \partial G_{23}-\bar \partial G_{13}\right)-2 \pi i  l\cdot \xi_3\right) \nn \\
	 &-2 \xi _1\cdot \xi _2 \bar \partial^2G_{12} \left(\xi _3\cdot k_1 \left(\bar \partial G_{23}-\bar \partial G_{13}\right)-2 \pi i  l\cdot \xi_3\right) \nn \\
	 &-2 \xi _1\cdot \xi _3 \bar \partial^2 G_{13} \left(\xi _2\cdot k_3 \left(\bar \partial G_{12}+\bar\partial G_{23}\right) -2 \pi i  l\cdot \xi _2\right)
	 \nn \\
	 &-2 \xi _2\cdot \xi _3 \bar \partial^2G_{23} \left(\xi _1\cdot k_2 \left(\bar \partial G_{12}-\bar \partial G_{13}\right)-2 \pi i  l\cdot \xi _1\right),
	\\
	\label{eq:lambda-2}	
	\hspace*{-9mm} \Lambda^2(\alpha_R,\beta_R,\bar z) = & 
\alpha'\Big[+(k_1\cdot k_2 \xi _1\cdot \xi _2 +\xi _1\cdot k_2 \xi _2\cdot k_3) \xi _3\cdot k_1 S^2_{\alpha_R\beta_R}\left(\bar z_{12}\right) \bar \partial G_{13}\nn \\
&-(k_1\cdot k_2 \xi _1\cdot \xi _2  +\xi _1\cdot k_2 \xi _2\cdot k_3) \xi _3\cdot k_1 S^2_{\alpha_R\beta_R}\left(\bar z_{12}\right) \bar \partial G_{23}\nn \\
&-(k_1\cdot k_3 \xi _1\cdot \xi _3 +\xi _1\cdot k_2  \xi _3\cdot k_1)\xi _2\cdot k_3 S^2_{\alpha_R\beta_R}\left(\bar z_{13}\right) \bar \partial G_{12} \nn \\
&-(k_2\cdot k_3 \xi _2\cdot \xi _3 +\xi _2\cdot k_3 \xi _3\cdot k_1) \xi _1\cdot k_2 S^2_{\alpha_R\beta_R}\left(\bar z_{23}\right) \bar \partial G_{12} \nn \\
&+(k_2\cdot k_3 \xi _2\cdot \xi _3 +\xi _2\cdot k_3 \xi_3\cdot k_1)\xi _1\cdot k_2 S^2_{\alpha_R\beta_R}\left(\bar z_{23}\right) \bar \partial G_{13} \nn \\
&-(k_1\cdot k_3 \xi _1\cdot \xi _3 +\xi _1\cdot k_2\xi _3\cdot k_1) \xi _2\cdot k_3 S^2_{\alpha_R\beta_R}\left(\bar z_{13}\right) \bar \partial G_{23} \nn \\
&+i2 \pi  \left(k_1\cdot k_2 \xi _1\cdot \xi _2+\xi _1\cdot k_2 \xi _2\cdot k_3\right) l\cdot \xi _3 S^2_{\alpha_R\beta_R}\left(\bar z_{12}\right)
\nn \\
&+i2 \pi  \left(k_1\cdot k_3 \xi _1\cdot \xi _3
+\xi _1\cdot k_2 \xi _3\cdot k_1\right) l\cdot \xi _2 S^2_{\alpha_R\beta_R}\left(\bar z_{13}\right) \nn \\
&+i2 \pi  (k_2\cdot k_3 \xi _2\cdot \xi _3  +2 \pi  \xi _2\cdot k_3 \xi _3\cdot k_1) l\cdot \xi _1 S^2_{\alpha_R\beta_R}\left(\bar z_{23}\right)\Big], \\
\label{eq:lambda-3}
\Lambda^3(\alpha_R,\beta_R,\bar z) &= \alpha' S_{\alpha_R\beta_R}\left(\bar z_{12}\right) S_{\alpha_R\beta_R}\left(\bar z_{13}\right) S_{\alpha_R\beta_R}\left(\bar z_{23}\right) \nn \\
&
(+2 \alpha'  \xi _1\cdot k_2 \xi _2\cdot k_3 \xi _3\cdot k_1 +\alpha  k_1\cdot k_2 \xi _1\cdot \xi _2 \xi _3\cdot k_1  \nn \\
&+\alpha'  k_1\cdot k_3 \xi _1\cdot \xi _3 \xi _2\cdot k_3 +\alpha  k_2\cdot k_3 \xi _2\cdot \xi _3 \xi _1\cdot k_2
).
\end{align}	
The quantity $\Phi_{\alpha_L^{}\beta_L^{}}^{}$ is given by
\begin{align}
	\Phi_{\alpha_L\beta_L}^{}(T^{a_1},T^{a_2},T^{a_3})
	= S_{\alpha_L\beta_L}^{}(z_{12}^{}) S_{\alpha_L\beta_L}^{}(z_{31}^{}) S_{\alpha_L\beta_L}^{}(z_{23}^{}) 
	Tr[T^{a_1}T^{a_2}T^{a_3}] \hs,
\end{align}
where $G(z)=-\log(\theta_1(z,\tau))$ and $S_{\alpha,\beta}(z)$ are the chiral scalar and fermion propagators 
on the torus respectively.\ In the above, we use the abbreviations 
$\bar{\partial}(\cdots\hsm)\!= \!{\partial(\cdots\hsm)}\hsm /{\partial\bar{z}}\hs$ and $G(z_{ij})=G_{ij}$.\ 
Here $\Lambda^1(\bar z)$ comes from $\langle \prod_{i} \xi_i \cdot \bar \partial X e^{i k_i \cdot X} \rangle$, which vanishes after spin-structure summation. Nevertheless, it is important to include $\Lambda^1(\bar z)$ in order to correctly decompose the scattering integrand and realize the double copy of internal loop states. 
 
The exact partition functions are given by
\begin{align}
	P_R(\alpha_R,\beta_R,\bar \tau) &= \frac{1}{2}\rho_{\alpha_R,\beta_R} \frac{\theta^4\Big[\begin{aligned}
			\alpha_R \\ \beta_R
		\end{aligned}\Big ](\bar z,\bar \tau)}{\eta^{12}(\bar \tau)}, \\
	P_L(\alpha_L,\beta_L,\tau)&=\frac{1}{2} \frac{\theta^{16}\Big[\begin{aligned}
			\alpha_L \\ \beta_L
		\end{aligned}\Big]( z, \tau)}{\eta^{24}(\tau)}, \\
	Z_{T^6}(\tau,\bar \tau) &=\prod_{i=1}^{6} \sum_{n,w} e^{-\pi \tau_2 (\frac{\alpha' n_i^2}{R_i^2}+\frac{w_i^2 R_i^2}{\alpha'})+2\pi i n w\tau_1}. 
\end{align}
Here $\rho_{\alpha_R,\beta_R}$ realizes the GSO projection with $\rho(0,0)=1$ and $\rho(0,1/2)=\rho(1/2,0)=-1$.

However, the spin-structure summation over the right-moving sector (summarized in more detail in Section \ref{sec:spin-sum}) gives 
\begin{align}
	\sum_{\alpha_R,\beta_R} P_R(\alpha_R,\beta_R) S^n_{\alpha_R \beta_R}(z,\tau)&=0,n=0,1,2,3,
\end{align}
showing that this one-loop three-point amplitude vanishes identically and hence gives a zero beta function. Counting the degrees of freedom, we obtain 
\begin{equation}
	c_{SYM_{N=4}}=c_A^{}+4c_F^{}+6c_S^{}=0 \hs.
\end{equation}
This cancellation corresponds to the vanishing of the beta function in $N=4$
SYM, consistent with its conformal invariance.

To explore the gravitational beta function in perturbation theory\footnote{Here the stabilization of the dilaton vacuum is not considered.}, we study the one-loop amplitude of three gravitons.  
The three-graviton one-loop amplitude is given by
\begin{align}
	\label{eq:A2h1-1oop}
	\hspace*{-3mm}
	\mathcal M_{3(1)}^{} \!=&\!\! \frac{\hat \kappa^3}{(2\pi)^3V}
	\!\sum_{\alpha_L,\beta_L}\!\sum_{\alpha_R,\beta_R}\!\!\int_{\mathcal F}^{}\!\!\! 
	\frac{\,\di^2\tau\hs\di^2\hsm z_2^{}\hs\di^2\hsm z_3^{}\,}{(4\pi^2)^2} \int d^4 l \nn \\ 
	&P_L(\alpha_L,\beta_L,\tau)
	P_R(\alpha_R,\beta_R,\bar \tau) Z_{T^6}
	\langle V_1^{1,2} U_2^{1,2} U_3^{1,2}\rangle_{[\alpha,\beta]}^{} \hs. 
\end{align}

In this way, the torus correlation function $\langle V^{(1,2)}_1 U^{(1,2)}_2 U^{(1,2)}_3 \rangle$ can be written in double-copy form
\begin{align}
	\label{eq:U1U2U3-g} 
	\langle V^{(1,2)}_1 U^{(1,2)}_2 U^{(1,2)}_3 \rangle_{[\alpha,\beta]}^{} 
	=&\frac{\alpha'^4}{64} \Big[(\Lambda^1(\bar z)+\Lambda^2(\alpha_R,\beta_R,\bar z)+\Lambda^3(\alpha_R,\beta_R,\bar z))\Lambda^b(z)
	\Big]. 
\end{align}
Here we define
\begin{align}
\label{eq:lambda-b}
\hspace*{-9mm}
\Lambda^b(z) =&  
\alpha' (\zeta_1\cdot k_2 ( \partial G_{12}- \partial G_{13})+2 \pi i  l\cdot \zeta_1) \nn \\
&(\zeta_2\cdot k_3 (\partial G_{12}+\partial G_{23})+2 \pi i  l\cdot \zeta_2) \nn \\
&\left(\zeta_3\cdot k_1 \left( \partial G_{23}- \partial G_{13}\right)+2 \pi i  l\cdot \zeta_3\right) \nn \\
&-2 \zeta _1\cdot \zeta_2  \partial^2G_{12} \left(\zeta_3\cdot k_1 \left(\partial G_{23}-\partial G_{13}\right)+2 \pi i  l\cdot \zeta_3\right) \nn \\
&-2 \zeta_1\cdot \zeta_3  \partial^2 G_{13} \left(\zeta_2\cdot k_3 \left(\partial G_{12}+\partial G_{23}\right) +2 \pi i  l\cdot \zeta_2\right)
\nn \\
&-2 \zeta_2\cdot \zeta_3  \partial^2G_{23} \left(\zeta_1\cdot k_2 \left(\partial G_{12}-\partial G_{13}\right)+2 \pi i  l\cdot \zeta_1\right).
\end{align}
It is clear that $\Lambda^b(z)$ is the complex conjugate of $\Lambda^1(\bar z)$ after replacing $\xi_i$ by $\zeta_i$.

\subsection{Decomposition of moduli space}
\label{sec:string-d-moduli}
The field-theory limit is realized by taking the world-sheet torus modulus to its world-line limit, $\tau_2^{} \!\ito\! \infty$.\ 
This corresponds to shrinking the radius of the closed string 
or forcing the closed string to propagate over an infinite distance. More details about the pinching rules were reviewed in Refs.~\cite{Balli:2024wje,Tourkine:2013rda,Tourkine:2019ukp}.\ 
After taking the world-line limit $\tau_2^{}\ito\infty$, there are two regions 
for each $z_{ij}^{}$ that correspond to Feynman diagrams in the world-line limit, 
as shown in Fig.~\ref{fig:3p1l}\footnote{For more general string amplitudes, more regions should be considered.}:
\begin{enumerate}
	\item[$\bullet$] 
	$|z_i^{}\!-\!z_j^{}|\!\sim\!\frac{1}{\,\tau_2^{}\,}$, shown as diagram (a) in Fig.~\ref{fig:3p1l}.
	\item[$\bullet$]  
	$|z_i^{}\!-\!z_j^{}|\!\sim\!\tau_2^{}$, shown as diagram (b) in Fig.~\ref{fig:3p1l}.
	
\end{enumerate}
We will refer to these regions as pinched diagrams and non-pinched diagrams, 
respectively. Different regions of $z_{ij}$ affect how $G(z_{ij},\bar z_{ij},\tau)$ and $S_{\alpha,\beta}(z_{ij},\tau)$ behave under the world-line limit $\tau_2\rightarrow +\infty$.
\begin{figure}
	\centering
	\subfloat{
		\begin{tikzpicture}[x=0.75pt,y=0.75pt,yscale=1,xscale=1]
			\draw [line width =1.5pt,color=blue] (0,0) circle (25);
			\draw [line width =1.5pt,color=blue] (50,8)--(40,8) arc (10:170:40)--(-60,8);
			\draw [line width =1.5pt,color=blue,rotate around x=180] (50,8)--(40,8) arc (10:170:40)--(-60,8);
			\draw [line width =1.5pt,color=blue,xscale=0.5] (100,0.2) circle (8);
			\draw [line width =1.5pt,color=blue,xscale=0.5] (-200,30) circle (8);
			\draw [line width =1.5pt,color=blue,xscale=0.5] (-200,-30) circle (8);
			\draw [line width =1.5pt,color=blue] (-100,38)--(-60,8);
			\draw [line width =1.5pt,color=blue,rotate around x=180] (-100,38)--(-60,8);
			\draw [line width=1.5pt,color=blue] (-100,22)--(-70,0);
			\draw [line width=1.5pt,color=blue] (-100,-22)--(-70,0);
			\coordinate [label=0:$1$] (1) at (-120,30);
			\coordinate [label=0:$2$] (2) at (-120,-30);
			\coordinate [label=0:$3$] (3) at (55,0);
			\coordinate [label=0:$\frac{1}{s_{12}}$] (4) at (-65,20);
			\coordinate [label=0:$(a)$] (3) at (-15,-80);
		\end{tikzpicture}
	}		
	\subfloat{
		\begin{tikzpicture}[x=0.75pt,y=0.75pt,yscale=1,xscale=1]
			\draw [line width =1.5pt,color=blue] (0,0) circle (25);
			\draw [line width =1.5pt,color=blue,xscale=0.5] (100,0.2) circle (8);
			\draw [line width =1.5pt,color=blue] (50,8)--(40,8) arc (10:110:40)--(-20,50);
			\draw [line width =1.5pt,color=blue,rotate around={120:(-27,46)}] (-27,46) ellipse (4 and 8);
			\draw [line width =1.5pt,color=blue,rotate around x=180] (50,8)--(40,8) arc (10:110:40)--(-20,50);
			\draw [line width =1.5pt,color=blue,rotate around={60:(-27,-46)}] (-27,-46) ellipse (4 and 8);
			\draw [line width=1.5pt,color=blue] (-35,-43)--(-27,-32) arc (231:130:40)--(-35,42);
			\coordinate [label=0:$1$] (1) at (-50,50);
			\coordinate [label=0:$2$] (2) at (-50,-50);
			\coordinate [label=0:$3$] (3) at (55,0);
			\coordinate [label=0:$(b)$] (3) at (-15,-80);
		\end{tikzpicture}	
	} 
	\vspace*{-3mm}
	\caption{\small %
		One-loop contributions to the three-point amplitude of closed strings.\ 
		For diagram (a), the distance between $z_1^{}$ and $z_2^{}$ is infinitesimally small, $\sim \frac{1}{\tau_2}$, as $\tau_2\ito \infty$ 
		and forms a propagator $\frac{1}{\,s_{12}^{}\,}$, whereas $z_3^{}$ is far from $z_1^{}$ and $z_2^{}$.\ 
		There are two similar diagrams with $z_{23}^{}\ito 0$ and  $z_{31}^{}\ito 0$ (forming the pole structures
		$\frac{1}{\,s_{23}^{}\,}$ and  $\frac{1}{\,s_{31}^{}\,}$) respectively.\ 
		For diagram (b), all $\{z_1^{},z_2^{},z_3^{}\}$ are far from one another. 
	}
	\label{fig:3p1l}
	\label{fig:1}
\end{figure}

Recall that the bosonic chiral propagator $G(z,\tau)$ is given by
\begin{equation}
	G(z,\tau)= -\log \theta_1(z,\tau). 
\end{equation}
Here we use the conventional abbreviation where $1$ denotes spin structure $(1/2,1/2)$, $2$ denotes spin structure $(1/2,0)$, $3$ denotes $(0,0)$ and $4$ denotes $(0,1/2)$. In the non-pinched regions, 
\begin{equation}
	G(z) =-\log(\sin(\pi z)) -4\sin^2(\pi z)q^2 + \mathcal{O}(q^4).
\end{equation}
Thus $W_0$ can be evaluated according to the following world-line limit $\tau_2\rightarrow \infty$ and $z\rightarrow i \tau_2$,
\begin{align}
	G(z,\tau)|_{q^0} \simeq & i\pi z\Theta(y),  \\
	\partial G(z,\tau)|_{q^0} \simeq &  i\pi \Theta(y), \\
	\partial^2 G(z,\tau)|_{q^0} \simeq &  \pi \delta(y). 
\end{align}
Here we count $\tau_2$ and $z=iy\tau_2$ as $\mathcal O(q^0)$.
The coefficient $W_2$ can be evaluated using
\begin{align}
	G(z)|_{q^2}&=-4\sin^2(\pi z)=e^{2i\pi z}+e^{-2i\pi z}-2, \\
	\partial G(z)|_{q^2} &=-4\pi \sin(2\pi z)=2i(e^{2\pi i z}-e^{-2\pi i z}), \\
	\partial^2 G(z)|_{q^2} &=-8\pi^2 \cos(2\pi z)=-4\pi^2(e^{2\pi i z}+e^{-2\pi i z}).
\end{align}

Suppose that $y=\Im z >0$, then $e^{2\pi i z} \rightarrow 0$ while $e^{-2\pi i z} \rightarrow \infty$ as $y\rightarrow \tau_2$. We can simply ignore $e^{2\pi i z}$ but $e^{-2\pi i z}$ must be handled carefully. It is worth noting that
\begin{align}
	e^{-2\pi i z}&= e^{2\pi y- 2\pi i x},\\
	e^{2\pi i \bz}&= e^{2\pi y+2\pi i x}.
\end{align}
We also need to integrate over $x$. Thus $e^{-2\pi i z}$ makes a nonzero contribution if and only if it is accompanied by $e^{2\pi i \bz}$. One can understand this factor by embedding it in a two-point amplitude, giving 
\begin{equation}
	\int_0^{\tau_2} dy e^{4\pi y} e^{-s\alpha'|y-y^2/\tau_2|} \sim \frac{1}{s-\frac{4}{\alpha'}}.
\end{equation}
Thus $e^{4\pi y}$ corresponds to exchanging a tachyon on another internal line. Since we are interested in massless exchanged states, and the tachyon must be canceled after combining the anti-holomorphic part, this factor $e^{-2\pi i z+ 2\pi i \bz}=e^{4\pi y}$ can be ignored.

As a consequence, we have the following effective replacement in the non-pinched region,
\begin{align}
	\label{eq:gzq2}
	G(z)|_{q^2}&\sim -2, \\
	\label{eq:p1gzq2}
	\partial G(z)|_{q^2} &\sim 0, \\
	\label{eq:p2gzq2}
	\partial^2 G(z)|_{q^2} &\sim 0.
\end{align}

On the other hand, in the pinched region $|z|\sim \frac{1}{\tau_2}$, the pole in $\log \theta_1(z,\tau)$ dominates and 
\begin{align}
	G(z,\bar z,\tau) & \simeq \log z, \\
	\partial G(z,\tau) & \sim \frac{1}{z}.
\end{align}
This is the same behavior as the scalar propagator on the complex plane. 
However, this scaling does not imply that the amplitude diverges. We change variables in the moduli integral from $d^2z =dx dy$ to polar coordinates $\rho d\rho d\theta$, where $\rho=|z|$ and $\theta\in [0,2\pi]$. Substituting, we obtain
\begin{equation}
	\int_0^l \rho d\rho d\theta \frac{1}{\rho^2}\rho^{\alpha'X} = \frac{2\pi l^{\alpha' X}}{\alpha'X}= \frac{2\pi }{\alpha' X}.
\end{equation}
where $X=(\sum_{i\in s} p_i)^2$ for the particles in the pinching set $s$, $l$ is an order-one free parameter, and in the second equation we take the limit $\alpha'\rightarrow 0$.

\subsection{Gauge beta function}
\label{sec:gauge-beta}
\subsubsection{Non-pinched diagram}
According to the analysis in Section \ref{sec:ILDC}, the contribution of the massless gauge sector to the one-loop three-gluon amplitude is given by

\begin{align}
	\hspace*{-3mm}
	\mathcal A_{3(1)} =\frac{~\hs\gh^3\! }{\,(2\pi)^3V\,} \int_{\mathcal F}^{}\!\!\! 
	\frac{\,\di^2\tau\hs\di^2\hsm z_2^{}\hs\di^2\hsm z_3^{}\,}{(4\pi^2  )^2} \int d^4 l\frac{\alpha'^2}{8}  (I_R^{\text{vector}}+I_{R}^{\text{scalar}}+I_{R}^{\text{fermion}})I_L^{\text{color}}. 
\end{align}

In the non-pinched region, after gauge fixing $z_1=0$, there are two distinct orderings: $0=y_1< y_2 < y_3<\tau_2$ and $0=y_1<y_3 < y_2<\tau_2$. Without loss of generality, we focus on the ordering $0=y_1<y_2 < y_3$ in this subsection; the ordering $0=y_1<y_3 < y_2$ can be evaluated in the same way. 

Thus for the $0=y_1<y_2 < y_3$ order,
\begin{align}
	\label{eq:lambda-1-np}
	\hspace*{-9mm}
	\Lambda^1(\bar z) =&  
          \alpha' (\xi_1\cdot k_2 (2\pi i)-2 \pi i  l\cdot \xi_1) (\xi_2\cdot k_3 (2\pi i)-2 \pi i  l\cdot \xi_2)\nn \\
          & \left(\xi_3\cdot k_1 \left(2\pi i\right)-2 \pi i  l\cdot \xi_3\right), 
	\\
	\label{eq:lambda-2-np}	
	\hspace*{-9mm} \Lambda^2(\alpha_R,\beta_R,\bar z) = & 
	\alpha'\Big[-(k_1\cdot k_2 \xi _1\cdot \xi _2 +\xi _1\cdot k_2 \xi _2\cdot k_3) \xi _3\cdot k_1 S^2_{\alpha_R\beta_R}\left(\bar z_{12}\right) \pi i\nn \\
	&-(k_1\cdot k_2 \xi _1\cdot \xi _2  +\xi _1\cdot k_2 \xi _2\cdot k_3) \xi _3\cdot k_1 S^2_{\alpha_R\beta_R}\left(\bar z_{12}\right) \pi i\nn \\
	&-(k_1\cdot k_3 \xi _1\cdot \xi _3 +\xi _1\cdot k_2  \xi _3\cdot k_1)\xi _2\cdot k_3 S^2_{\alpha_R\beta_R}\left(\bar z_{13}\right) \pi i \nn \\
	&-(k_2\cdot k_3 \xi _2\cdot \xi _3 +\xi _2\cdot k_3 \xi _3\cdot k_1) \xi _1\cdot k_2 S^2_{\alpha_R\beta_R}\left(\bar z_{23}\right) \pi i \nn \\
	&-(k_2\cdot k_3 \xi _2\cdot \xi _3 +\xi _2\cdot k_3 \xi_3\cdot k_1)\xi _1\cdot k_2 S^2_{\alpha_R\beta_R}\left(\bar z_{23}\right) \pi i \nn \\
	&-(k_1\cdot k_3 \xi _1\cdot \xi _3 +\xi _1\cdot k_2\xi_3\cdot k_1) \xi _2\cdot k_3 S^2_{\alpha_R\beta_R}\left(\bar z_{13}\right) \pi i \nn \\
	&+i2 \pi  \left(k_1\cdot k_2 \xi _1\cdot \xi _2+\xi _1\cdot k_2 \xi _2\cdot k_3\right) l\cdot \xi _3 S^2_{\alpha_R\beta_R}\left(\bar z_{12}\right)
	\nn \\
	&+i2 \pi  \left(k_1\cdot k_3 \xi _1\cdot \xi _3
	+\xi _1\cdot k_2 \xi _3\cdot k_1\right) l\cdot \xi _2 S^2_{\alpha_R\beta_R}\left(\bar z_{13}\right) \nn \\
	&+i2 \pi  (k_2\cdot k_3 \xi _2\cdot \xi _3  +2 \pi  \xi _2\cdot k_3 \xi _3\cdot k_1) l\cdot \xi _1 S^2_{\alpha_R\beta_R}\left(\bar z_{23}\right)\Big], \\
	\label{eq:lambda-3-np}
	\Lambda^3(\alpha_R,\beta_R,\bar z) &= \alpha' S_{\alpha_R\beta_R}\left(\bar z_{12}\right) S_{\alpha_R\beta_R}\left(\bar z_{13}\right) S_{\alpha_R\beta_R}\left(\bar z_{23}\right) \nn \\
	&
	(+2 \alpha'  \xi _1\cdot k_2 \xi _2\cdot k_3 \xi _3\cdot k_1 +\alpha  k_1\cdot k_2 \xi _1\cdot \xi _2 \xi _3\cdot k_1  \nn \\
	&+\alpha'  k_1\cdot k_3 \xi _1\cdot \xi _3 \xi _2\cdot k_3 +\alpha  k_2\cdot k_3 \xi _2\cdot \xi _3 \xi _1.k_2
	)
\end{align}
and 
\begin{align}
S^2_{0,0}(\bar z_{ij})& \simeq 0\times q^0 - 8\pi^2 q + 16\pi^2 q^2, \\
S^2_{1/2,0}(\bar z_{ij})& \simeq -\pi^2 + 0\times q +0\times q^2, \\	
S_{0,0}\left(\bar z_{12}\right) S_{0,0}\left(\bar z_{31}\right) S_{0,0}\left(\bar z_{23}\right) & \simeq 0\times q^0 - 8\pi^3 i\times q^1  + 0\times q^2 + ... \\
S_{1/2,0}\left(\bar z_{12}\right) S_{1/2,0}\left(\bar z_{31}\right) S_{1/2,0}\left(\bar z_{23}\right) & \simeq -i\pi^3\times q^0 + 0\times q^1  + 0\times q^2 + ...
\end{align}

Before evaluating this integral, we observe that not all terms in the above equations contribute to the renormalization of the tree-level three-gluon vertex. However, it is nontrivial to determine whether the $ l $-dependent terms affect the physical running. Therefore, it is useful to study the typical integrals over the loop momentum and the fundamental region of $ \tau $.

We observe that in the non-pinched region,
\begin{align}
 |\mathcal{J}_3|^2 &= e^{-\alpha'\pi \tau_2 l^2 - 2\pi \alpha' \sum_i l\cdot k_i y_i + \pi \alpha' \sum_{i<j} k_i\cdot k_j |y_{ij}|} + \mathcal{O}(q^2) \nn \\
 &=e^{-\pi \alpha' \tau_2 p^2 + \pi \alpha' \tau_2 (\sum_i k_i \hat y_i)^2+\pi \alpha' \tau_2 \sum_{i<j} k_i\cdot k_j |\hat y_{ij}| } + \mathcal{O}(q^2),
\end{align}
where we define $p=l+\sum_i k_i y_i$ and $\hat y_i=\frac{y_i}{\tau_2}$. For later convenience, we will denote $\tilde k=\sum_a k_a y_a$, $\Delta=\pi \alpha' \tau_2 (\sum_i k_i \hat y_i)^2+\pi \alpha'\sum_{i<j} k_i\cdot k_j |\hat y_{ij}|$ and
\begin{equation}
	J_3=|\mathcal{J}_3|=e^{-\frac{\pi}{2} \alpha' \tau_2 p^2 + \frac{\pi}{2} \alpha' \tau_2 (\sum_i k_i \hat y_i)^2+\frac{\pi}{2} \alpha' \tau_2 \sum_{i<j} k_i\cdot k_j |\hat y_{ij}| }.
\end{equation}
Recall that 
\begin{equation}
	(\sum_i k_i \hat y_i)^2=2\sum_{i<j} k_i.k_j\hat y_i \hat y_j =-\sum_{i<j}k_i.k_j \hat y_{ij}^2,
\end{equation}
where we used the fact that $\sum_{i\neq j}k_i \cdot k_j =0$. 
Thus the standard result $\Delta=\pi \alpha' k_i.k_j\sum_{i<j}(|\hat y_{ij}|-\hat y_{ij}^2)$ in the conventional RNS formalism is obtained.

A typical integral can then be expressed as 
\begin{equation}
	I(n,m)=\int_{\mathcal F} d^2\tau \int d^4 l l^{n} \tau_2^m e^{-\pi\alpha \tau_2 (l^2-\Delta)}.
\end{equation}
If one is interested only in the field-theory limit, $I(n,m)$ can be reduced to the world-line limit and regulated by conventional dimensional regularization,
\begin{align}
	I^1(n,m)&=\int_{0}^\infty d\tau_2 \int d^{4-\epsilon} l l^{n} \tau_2^m e^{-\pi\alpha \tau_2 (l^2-\Delta)} \nn \\
                &=\int d^{4-2\epsilon} l l^{n}\frac{m!}{[\pi \alpha'(l^2-\Delta)]^{m+1}}.
\end{align}
Comparing with the Feynman integral in quantum field theory, we see that $\{\hat y_i\}$ play the same role as Feynman parameters.

We also follow the alternative regularization procedure of Ref.~\cite{Green:2008uj} and introduce a new cutoff $L$ 
in the fundamental region as shown in Fig.\,\ref{fig:2}, 
which is used to extract the field-theory limit as $L\ito \infty\hs$.\ 
The physical result should be independent of $L$.\ 
Any $L$-dependent term obtained from the point-particle region $[L,\infty]$ of the integral 
will be canceled by another $L$-dependent term 
from the stringy region, marked in pink in Fig.\,\ref{fig:2}. The $L$-regulated integral is given by 
\begin{align}
	I^2(n,m)&=\int_{L}^\infty d\tau_2 \int d^{4} l l^{n} \tau_2^m e^{-\pi\alpha \tau_2 (l^2-\Delta)}.
\end{align}
A more detailed comparison between $L$-regularization and dimensional regularization can be found in Appendix \ref{sec:L-regulation}. 
Ultimately, these two regularization methods are demonstrated to be equivalent within the theoretical framework of renormalization.

\begin{figure}
	\centering
	\subfloat{
		\begin{tikzpicture}[x=0.75pt,y=0.75pt,yscale=1,xscale=1]
			\draw [color=blue!50,line width=38pt]
			(0,80)--(0,110);
			\draw [color=red!50,line width=38pt]
			(0,0)--(0,80);
			\filldraw [color=black!1,line width=1pt]
			(50,0) arc (0:180:50);
			\draw [color=black,line width=1pt,->]
			(-100,0)--(100,0);
			\draw [color=black,line width=1pt,->]
			(0,-20)--(0,120);
			\draw [color=black,line width=1pt]
			(50,0) arc (0:180:50);
			\draw [color=black,line width=1pt]
			(-25,43.3)--(-25,110);
			\draw [color=black,line width=1pt,dash dot dot]
			(-25,43.3)--(-25,0);
			\draw [color=black,line width=1pt]
			(25,43.3)--(25,110);
			\draw [color=black,line width=1pt,dash dot dot]
			(25,43.3)--(25,0);
			\coordinate [label=0:$L$] (L) at (0,90);
			\draw [color=black,line width=2pt,dash dot]
			(-50,80)--(50,80);
			\coordinate [label=0:$\tau_1$] (x) at (100,0);
			\coordinate [label=0:$\tau_2$] (y) at (-10,125);
			\coordinate [label=0:$1$] (y) at (42.5,-10);
			\coordinate [label=0:$0.5$] (y) at (12,-10);
			\coordinate [label=0:$-0.5$] (y) at (-45,-10);
			\coordinate [label=0:$-1$] (y) at (-65,-10);
	\end{tikzpicture} }		
	\vspace*{-2mm}	 	 
	\caption{\small %
		The fundamental region is the colored regions where $|\tau|\hsm\!>\hsm\!1$ and 
		$-\frac{1}{2}\!\!<\!\tau_1^{}\!\!<\!\frac{1}{2}\hs$.\   
		A cutoff $L\!\!\gg\!\! 1$ on the $\tau_2^{}$ axis is introduced for the following calculation.\ 
		The $\tau_2^{}\!>\!L$ region
		corresponds to the field-theory-limit region, and 
		$\tau_2^{} \hsm\!<\!L$ is the stringy region.}
	\label{fig:2}
\end{figure}

In addition, it should be stressed that both $I^1(n,m)$ 
and $I^2(n,m)$ 
converge for $l^2-\Delta>0$; otherwise these integrals diverge.
The traditional strategy is to evaluate the integral in the momentum region where convergence is guaranteed and then analytically continue the result to the divergent region. This procedure is universally valid in the low-energy limit $\alpha's_{ij} \rightarrow 0$. For finite $\alpha' s_{ij}$, 
however, contour deformation of the integral becomes necessary. The related framework of contour deformation was developed in Refs.~\cite{Witten:2013pra,Eberhardt:2023xck}.

In this way,
\begin{equation}
	I(0,2)\sim \frac{1}{\Delta},
\end{equation}
which leads to an IR divergence but no UV divergence. Moreover, $I(1,2)=I(3,2)=0$ because of the reversal symmetry $l\rightarrow -l$. Only
\begin{align}
	\label{eq:I22}
	I^1(2,2)&=\frac{2}{(\pi\alpha')^{3}\Gamma(3)}i\pi^{2-\epsilon}(2-\epsilon)(\frac{1}{\epsilon}-\gamma)(1-\epsilon\log \Delta) + \mathcal{O}(\epsilon^1)
\end{align}
provides a nonzero UV correction. As a consequence, only terms proportional to $p^2$ in $\Lambda^1(\bar z)$ make nonzero contributions in the non-pinched region. These terms are given by
\begin{equation}
\Lambda^1_{\text{red}}(\bar z)=\alpha' (2\pi i)^3 
\frac{1}{4-2\epsilon}l^2\xi_1.\xi_2
 \xi_3.k_1(1 +y_{12})+\text{cycle}. 
\end{equation}  
Here we use the effective replacement $l^\mu l^\nu \rightarrow \frac{1}{d}l^2 g^{\mu\nu}$ in the loop momentum integral. Thus
\begin{align}
	I_R^{\text{scalar}}\sim&n_{\text{s}} \Lambda^1_{\text{red}}(\bar z) J_3, \\
	I_R^{\text{vector}}\sim&(d-2)\Lambda^1_{\text{red}}(\bar z) J_3, \\
    I_R^{\text{fermion}}\sim &-2n_f5\Lambda^1_{\text{red}}(\bar z)	J_3.
\end{align} 

On the other hand, 
\begin{align}
I_L^{\text{color}}&\sim 4\pi^3 i C_{\text{SO}(32)} J_3. 
\end{align}
After subtracting the constant and divergent parts of $I^1(2,2)$. 
The renormalized non-pinched diagram is given by 
\begin{align}
\mathcal{A}^{\text{NP}}_{3(1)} =&\frac{~\hs\gh^3\! }{\,(2\pi)^3V\,}2\times \int_0^{\hat y_3} d\hat y_2 \int_0^1 d\hat y_3 \frac{\pi}{8}\xi_1\cdot \xi_2\xi_3\cdot k_1 (1+y_{12})
\nn \\
&(d-2+n_s-2n_f)(-\log (\Delta/\mu^2)) + \text{cycle}. 
\end{align}
Here the factor of $2$ accounts for the contribution from the alternative ordering $0=y_1<y_3<y_2$.
Evaluating the residue integral over the Feynman parameters $\hat y_2$ and $\hat y_3$ is highly nontrivial and beyond the scope of this paper. Detailed discussions can be found in the textbook \cite{Smirnov:2006ry}. Fortunately, since the beta function depends solely on the renormalization scale $\mu$, we focus on the scale-dependent part of 
$\mathcal{A}^{\text{NP}}_{3(1)}$ which is given by
\begin{equation}
	\mathcal{A}^{\text{NP}}_{3(1)}(\mu)=\frac{~\hs\gh^3\! }{\,(2\pi)^3V\,}\xi_1\cdot \xi_2 \xi_3\cdot k_{1} \frac{\pi}{2\times 3}(d-2+n_s-2n_f)\log \mu+ \text{cycle}. 
\end{equation}
This expression reveals the exact cancellation in supersymmetric theories, where $d-2+n_s-2n_f=0$. 
It explains why pinched diagrams alone suffice to compute the full beta function of the $N=1$ orbifolded heterotic string in Ref.~\cite{Minahan:1987ha}.

\subsubsection{Pinched diagram}
There are three pinched diagrams, or regions,
\begin{equation}
	\{|z_{12}|\sim \frac{1}{\tau_2},|z_{23}|\sim \frac{1}{\tau_2},|z_{31}|\sim \frac{1}{\tau_2}\}.
\end{equation}
In this subsection, we focus on the first region, $|z_{12}|\sim \frac{1}{\tau_2}$ and $|z_{23}|\sim \tau_2$. After gauge fixing $z_1=0$, $|z_{12}|\sim \frac{1}{\tau_2}$ allows two cases: $|z_2|\sim \frac{1}{\tau_2}$ and $|z_2-\tau_2|\sim \frac{1}{\tau_2}$. The second case arises from periodicity on the torus.  When $z_1$ and $z_2$ pinch, both $\partial G(z_{12})$ and $S_\delta(z_{12})$ scale as $\frac{1}{z_{12}}$ for any spin structure $\delta$. However, this scaling does not imply that the amplitude diverges. We change variables in the moduli integral from $d^2z_2=dx_2 dy_2$ to polar coordinates $\rho d\rho d\theta$, where $\rho=|z_2|$ and $\theta\in [0,2\pi]$. Substituting, we obtain
\begin{equation}
	A_{3(1)}^{\text{P}}\sim \int_0^L \rho d\rho d\theta \frac{1}{\rho^2}\rho^{\alpha'k_1\cdot k_2} = \frac{2\pi L^{\alpha' k_1 \cdot k_2}}{\alpha'k_1 \cdot k_2}= \frac{2\pi }{\alpha'k_1 \cdot k_2}.
\end{equation}
Here $L$ is an order-one free parameter, and in the second equation we take the limit $\alpha'\rightarrow 0$.
Thus, as shown in Fig.~\ref{fig:1}, the pinching creates a pole $\frac{2\pi}{\alpha' k_1 \cdot k_2}$. Therefore, we only need to consider the residue at $z_{ij}\rightarrow 0$.

In this way, it is easy to evaluate the residues at $z_{12}\rightarrow 0$  as 
\begin{align}
	\text{Res}_{z_{12}} \Phi(\alpha,\beta,z)&= S_{\alpha\beta}\hsm ( z_{13}^{}) S_{\alpha\beta}^{}\hsm( z_{31}^{}),
	\\
	\label{eq:plambda-1}
	\hspace*{-9mm}
	\text{Res}_{\bar z_{12}} \Lambda^1(\bar z) =&  
	\alpha' (\xi_1\cdot k_2) (\xi_2\cdot k_3 (\bar \partial G_{23})-2 \pi i  l\cdot \xi_2) \left(-2 \pi i  l\cdot \xi_3\right) \nn \\
	&+\alpha' (\xi_1\cdot k_2 (\bar \partial G_{31})-2 \pi i  l\cdot \xi_1) (\xi_2\cdot k_3) \left(-2 \pi i  l\cdot \xi_3\right),
	\\
	\label{eq:plambda-2}	
	\hspace*{-9mm} \text{Res}_{\bar z_{12}}\Lambda^2(\alpha_R,\beta_R,\bar z) = & 
	\alpha'\Big[ 
	-(k_1\cdot k_3 \xi _1\cdot \xi _3 +\xi _1\cdot k_2  \xi _3\cdot k_1)\xi _2\cdot k_3 S^2_{\alpha_R\beta_R}\left(\bar z_{13}\right)  \nn \\
	&-(k_2\cdot k_3 \xi _2\cdot \xi _3 +\xi _2\cdot k_3 \xi _3\cdot k_1) \xi _1\cdot k_2 S^2_{\alpha_R\beta_R}\left(\bar z_{23}\right)  \Big], \\
	\label{eq:plambda-3}
	\text{Res}_{\bar z_{12}}\Lambda^3(\alpha_R,\beta_R,\bar z) &= \alpha'  S_{\alpha_R\beta_R}\left(\bar z_{13}\right) S_{\alpha_R\beta_R}\left(\bar z_{23}\right) \nn \\
	&
	(+2 \alpha'  \xi _1\cdot k_2 \xi _2\cdot k_3 \xi _3\cdot k_1 +\alpha  k_1\cdot k_2 \xi _1\cdot \xi _2 \xi _3\cdot k_1  \nn \\
	&+\alpha'  k_1\cdot k_3 \xi _1\cdot \xi _3 \xi _2\cdot k_3 +\alpha  k_2\cdot k_3 \xi _2\cdot \xi _3 \xi _1\cdot k_2
	).
\end{align}	
We observe that some terms again contribute to the $\alpha' F^3$ effective vertex. After removing the terms that are irrelevant to the beta function, the reduced formulas are 
\begin{align}
	\label{eq:plambda-1-red}
	\hspace*{-9mm}
	&\text{Res}_{\bar z_{12}} \Lambda^1_{\text{red}}(\bar z) \nn \\  
	= &  
	\alpha' (\xi_1\cdot k_2) (2 \pi i  p\cdot \xi_2) \left(2 \pi i  p\cdot \xi_3\right) 
	+\alpha' (2 \pi i  p\cdot \xi_1) (\xi_2\cdot k_3) \left(2 \pi i  p\xi_3\right),
	\\
	\label{eq:plambda-2-red}
	\hspace*{-9mm} &\text{Res}_{\bar z_{12}}\Lambda^2_{\text{red}}(\alpha_R,\beta_R,\bar z) +\text{Res}_{\bar z_{12}}\Lambda^3_{\text{red}}(\alpha_R,\beta_R,\bar z) \nn \\ 
 =& \alpha'  S_{\alpha_R\beta_R}\left(\bar z_{13}\right) S_{\alpha_R\beta_R}\left(\bar z_{23}\right) (\alpha  k_1\cdot k_2 \xi _1\cdot \xi _2 \xi _3\cdot k_1 ).
\end{align}

Once $|z_{12}|\sim \frac{1}{\tau_2}$ and $z_1=0$, the Koba-Nielsen factor reduces to 
\begin{align}
	\lim_{z_{12}\rightarrow 0}|\mathcal{J}_3|^2 
	&=e^{-\pi \alpha' \tau_2 (l^2-\Delta'_{12})  }  + \mathcal{O}(q^2), \\
	\Delta'_{12}&=\lim_{z_{12}\rightarrow 0} \Delta = \pi \alpha' k_1\cdot k_2(y_3^2-|y_3|).
\end{align} 

Thus we have
\begin{align}
	\text{Res}_{z_{12}}I_L^{\text{color}}\sim & 4\pi^2 C_{\text{SO}(32)} J_3^{(12)},\\
	\text{Res}_{z_{12}}I_R^{\text{scalar}}\sim &  
	-4\pi^2 \alpha'n_s \frac{l^2}{4-2\epsilon} (\xi_2\cdot \xi_3\xi_1\cdot k_2  + \xi_1\cdot \xi_3\xi_2\cdot k_3)J_3^{(12)},\\
	\text{Res}_{z_{12}}I_R^{\text{vector}}\sim & -4\pi^2 \alpha'(d-2) \frac{l^2}{4-2\epsilon} (\xi_2\cdot \xi_3\xi_1\cdot k_2  + \xi_1\cdot \xi_3\xi_2\cdot k_3)J_3^{(12)} \nn\\
	& -\alpha' 8\pi^2 (k_1\cdot k_2 \xi _1\cdot \xi _2 \xi _3\cdot k_1 )J_3^{(12)}, \\                
	\text{Res}_{z_{12}}I_R^{\text{fermion}}\sim&
	8\pi^2 \alpha' n_f \frac{l^2}{4-2\epsilon} (\xi_2\cdot \xi_3\xi_1\cdot k_2  + \xi_1\cdot \xi_3\xi_2\cdot k_3)J_3^{(12)}
	\\
	&+\alpha'^2 2 \pi^2 n_f (k_1\cdot k_2 \xi _1\cdot \xi _2 \xi _3\cdot k_1 )J_3^{(12)},
\end{align}
where $J_3^{(12)}=\lim_{z_{12}\rightarrow 0}|\mathcal{J}_3|$.
Hence, only two integrals are associated with the pinched diagrams,
\begin{align}
	\label{eq:I21}
	I(2,1) &= \frac{-i\Gamma(\epsilon-1)}{\alpha'^2 \pi^\epsilon \Gamma(2)}(2-\epsilon)(\Delta'_{12})^{1-\epsilon},  \\
	\label{eq:I01}
	I(0,1) &=\frac{i\Gamma(\epsilon)}{\alpha'^2\pi^\epsilon \Gamma(2)} (\frac{1}{\Delta'_{12}})^{\epsilon}.
\end{align}

In this way, one can check that the renormalization-scale-dependent part is given by
\begin{align}
	 A_{3(1)}(\mu)^{\text{P}}(|z_{12}\sim   
	\frac{1}{\tau_2}|,\mu) =& -\frac{\pi}{2}\frac{~\hs\gh^3\! }{\,(2\pi)^3V\,}\zeta_1 \cdot \zeta_2 \zeta_3 \cdot k_1 C_{SO(32)} \log(\mu)(4-n_f), \nn \\
	&-\frac{\pi}{2}\frac{~\hs\gh^3\! }{\,(2\pi)^3V\,} \zeta_2 \cdot \zeta_3 \zeta_1 \cdot k_2C_{SO(32)} \log(\mu)(\frac{d-2}{12}+\frac{n_s}{12}-\frac{2n_f}{12}), \nn \\
	&- \frac{\pi}{2}\frac{~\hs\gh^3\! }{\,(2\pi)^3V\,} \zeta_3 \cdot \zeta_1 \zeta_2 \cdot k_3 C_{SO(32)} \log(\mu)(\frac{d-2}{12}+\frac{n_s}{12}-\frac{2n_f}{12}).
\end{align}

Summing over the three pinched regions,

\begin{align}
A_{3(1)}^{\text{P}}(\mu) \sim  
&-\frac{\pi}{2}\frac{~\hs\gh^3\! }{\,(2\pi)^3V\,}C_{SO(32)}(K[1,2;3]+\text{cycle}) \log(\mu)(4-n_f+\frac{d-2}{6}+\frac{n_s}{6}-\frac{2n_f}{6}), 
\end{align}
where $K[1,2;3]\equiv \zeta_1 \cdot \zeta_2 \zeta_3 \cdot k_1$. 

Combining this with $A_{3(1)}^{\text{NP}}$, it is easy to verify that when $d=4$,
\begin{align}
		A_{3(1)}(\mu)  \sim -\frac{i \hat g^3}{16\pi^2 V} 
		& C_{SO(32)} (K[1,2;3]+\text{cycle}) \log(\mu)(\frac{11}{3}-\frac{2}{3}\times n_f -\frac{n_s}{6}). 
\end{align}
This corresponds to the beta function given in Eq.~\eqref{eq:beta-field}. Thus far, the decomposed string amplitude gives the same result as quantum field theory, and the result is almost model independent\footnote{The kinematic part is model independent, but the color part is nontrivial to generalize to an arbitrary Lie group.}.

\subsection{Gravitational beta function}
\label{sec:gravi-beta}
In this section, we shift our focus to the gravitational beta function. Owing to the loop-level double-copy structure inherent in the chiral-splitting formalism, it is sufficient to replace $\Phi_{\alpha_L^{}\beta_L^{}}$ with $\Lambda^B(z)$ derived in Section \ref{sec:ILDC}. Surprisingly, we need not specify how to split $W(z)=\Lambda^B(z)$ into components; the tensor structure of this integral already forces the beta function of the gravitational coupling to vanish.

Since $l\cdot \zeta_i$ factors are provided by $\Lambda^B_{\text{red}}$, the following effective replacements in the loop-momentum integrand should be taken into account,
\begin{align}
  \label{eq:replace-1}
  l\cdot \xi_i l\cdot \zeta_j &=\frac{l^2}{d}\xi_i\cdot \zeta_j, \\
  \label{eq:replace-2}
  l\cdot \xi_i l\cdot \xi_j l\cdot \zeta_n l\cdot \zeta_m &= \frac{l^4}{d(d+2)}(\xi_i\cdot \xi_j\zeta_n\cdot \zeta_m+\xi_i\cdot \zeta_n\xi_j\cdot \zeta_m+\xi_i\cdot \zeta_m\xi_j\cdot \zeta_n)
\end{align}
It is worth noting that there is no $\xi_i.\xi_j$ in Eq.~\eqref{eq:lambda-1-np} as well as $\Lambda^B(z)$ in the non-pinched region. Therefore,  
\begin{align}
(\Lambda^1(\bar z)\Lambda^B(z))_{\text{red}}=&[\alpha' (2\pi i)^3 
l\cdot \xi_1 l\cdot \xi_2
\xi_3\cdot k_1(1 +y_{12})+\text{cycle}]
\nn \\
&[\alpha' (-2\pi i)^3 
l.\zeta_1 l\cdot \zeta_2
\zeta_3\cdot k_1(1 +y_{12})+\text{cycle}].
\end{align}
Since the polarization tensors of physical gravitons are symmetric and traceless, we take $\xi_i\cdot \zeta_i=0$ in Eqs.~\eqref{eq:replace-1} and \eqref{eq:replace-2}. Similarly,
$\Lambda^2(\alpha_R,\beta_R,\bar z)$ and $\Lambda^3(\alpha_R,\beta_R,\bar z)$ either provide a $\xi_i\cdot \xi_j$, $l\cdot \xi_i l\cdot \xi_j$ or $\xi_i\cdot \xi_j l\cdot \xi_l$, 
\begin{align}
(\Lambda^2(\alpha_R,\beta_R,\bar z)\Lambda^B(z))_{\text{red}} = & 
	\Big[-2\pi i(\alpha 'k_1\cdot k_2) \xi _1\cdot \xi _2  \xi _3\cdot k_1(1+y_{12}) S^2_{\alpha_R\beta_R}\left(\bar z_{12}\right) \nn \\
	&-2\pi i (\alpha'k_1\cdot k_3) \xi _1\cdot \xi _3 \xi _2\cdot k_3 (1+y_{31}) S^2_{\alpha_R\beta_R}\left(\bar z_{13}\right) \nn \\
	&-2\pi i(\alpha'k_2\cdot k_3) \xi _2\cdot \xi _3  \xi _1\cdot k_2 (1+y_{23}) S^2_{\alpha_R\beta_R}\left(\bar z_{23}\right)  \Big]
	\nn \\
	&[\alpha' (-2\pi i)^3 
	l\cdot \zeta_1 l\cdot \zeta_2
	\zeta_3\cdot k_1(1 +y_{12})+\text{cycle}] \nn \\
	&+\Big[i2 \pi  (\alpha'k_1\cdot k_2) \xi _1\cdot \xi _2 l\cdot \xi _3 S^2_{\alpha_R\beta_R}\left(\bar z_{12}\right)
	\nn \\
	&+i2 \pi (\alpha'k_1\cdot k_3) \xi _1\cdot \xi _3
	l\cdot \xi _2 S^2_{\alpha_R\beta_R}\left(\bar z_{13}\right)
	 \nn \\
	&+i2 \pi  (\alpha'k_2\cdot k_3) \xi _2\cdot \xi _3 l\cdot \xi _1 S^2_{\alpha_R\beta_R}\left(\bar z_{23}\right)\Big]
	\nn \\
	&\Big[-(-2\pi i)^3 l\cdot \zeta_1 \zeta_2\cdot k_3 (1+y_{31}) \zeta_3\cdot k_1 (1+y_{12}) +\text{cycle} \Big],
	 \\
	(\Lambda^2(\alpha_R,\beta_R,\bar z)\Lambda^B(z))_{\text{red}} =&  S_{\alpha_R\beta_R}\left(\bar z_{12}\right) S_{\alpha_R\beta_R}\left(\bar z_{13}\right) S_{\alpha_R\beta_R}\left(\bar z_{23}\right) \nn \\
	&
	( \alpha  k_1\cdot k_2 \xi _1\cdot \xi _2 \xi _3\cdot k_1  +\alpha'  k_1\cdot k_3 \xi _1\cdot \xi _3 \xi _2\cdot k_3 +\alpha  k_2\cdot k_3 \xi _2\cdot \xi _3 \xi _1\cdot k_2
	) \nn \\
	&[\alpha' (-2\pi i)^3 
	l\cdot \zeta_1 l\cdot \zeta_2
	\zeta_3\cdot k_1(1 +y_{12})+\text{cycle}].
\end{align}
If these integrals do not vanish due to supersymmetry,
only two kinds of integrals are required, namely
\begin{align}
	I^1(4,2)&=\frac{-i2\pi^{2-\epsilon}}{\pi^3 \alpha'^3} (2-\epsilon)(3-\epsilon) \frac{\Gamma(\epsilon-1)}{2} \Delta^{1-\epsilon},   
\end{align}
and $I^1(2,2)$ as shown in Eq.~\eqref{eq:I22}. It is worth noting that the loop corrections in the non-pinched diagram are universally proportional to $s_{ij}\log (C\Delta)$ before integration over Feynman parameters, where $C$ is a constant depending on the regularization scheme. The most general renormalization condition can be chosen as  
\begin{equation}
	s_{12}=a_{12}\mu^2,s_{23}=a_{23}\mu^2,s_{31}=a_{31}\mu^2
\end{equation} 
with $a_{12}+a_{23}+a_{31}=0$ and $a_{ij}\in R$. However, the subtraction procedure remains ambiguous. Conventionally, the renormalized one-loop amplitude could be proportional to 
\begin{equation}
	s_{nm}\log(\Delta) - x s_{nm}\log(\sum_{i<j}\pi \alpha'(y^2_{ij}-|y_{ij}|)a_{ij}\mu^2) -y  a_{nm} \mu^2\log(\sum_{i<j}\pi \alpha'(y^2_{ij}-|y_{ij}|)a_{ij}\mu^2),
\end{equation}
for any $x+y=1$ and $x,y\in R$. Physically, $a_{nm} \mu^2\log(\sum_{i<j}\pi \alpha'(y^2_{ij}-|y_{ij}|)a_{ij}\mu^2)$ corresponds to power-law running of the gravitational coupling, while $s_{nm}\log(\sum_{i<j}\pi \alpha'(y^2_{ij}-|y_{ij}|)a_{ij}\mu^2)$ generates a loop correction to a higher-dimensional operator. 

However, the term $a_{nm} \mu^2\log(\sum_{i<j}\pi \alpha'(y^2_{ij}-|y_{ij}|)a_{ij}\mu^2)$ introduces dependence of the beta function on the physically irrelevant parameters $a_{ij}$. Thus the only self-consistent choice is $x=1$ and $y=0$, which ensures that the running of the physical coupling is independent of the artificial renormalization conditions. Beyond this renormalization argument, the result is also corroborated by 
the $L$-regularization 
\begin{equation}
\label{eq:I42-L}
 I^2(4,2) =\frac{\Gamma(5)}{(\pi\alpha')^4 \Gamma(2)}(\pi \Delta) \int d\alpha' (-\log L -\gamma -\log(\pi\alpha' \Delta)+\mathcal{O}(L^1)).
\end{equation}
Since the string-derived counterterms are in one-to-one correspondence with the $L$-dependent terms given by $L$-regularization, Eq.~\eqref{eq:I42-L} demonstrates that all the counterterms provided by string theory should be proportional to $s_{ij}$, which corresponds only to $s_{nm}\log (\sum_{i<j}\pi \alpha'\\ (y^2_{ij}-|y_{ij}|)a_{ij}\mu^2)$.  
Consequently, the non-pinched diagram contributes loop corrections to higher-dimensional operators but does not renormalize the gravitational coupling.

Next, without loss of generality, we consider the region $|z_{12}|\sim \frac{1}{\tau_2}$. The residue of $\Lambda^B_{\text{red}}(z)$ is given by 
\begin{equation}
	\text{Res}_{z_{12}} \Lambda^B_{\text{red}}(z) =  
	\alpha' (\xi_1\cdot k_2) (+2 \pi i  l\cdot \xi_2) \left(+2 \pi i  l\cdot \xi_3\right).
\end{equation}
With Eq.~\eqref{eq:plambda-1-red} and Eq.~\eqref{eq:plambda-2-red},
the relevant integrals are
\begin{align}
	I(4,1)&=\frac{i2\pi^{2-\epsilon}}{\pi^2 \alpha'^2} (2-\epsilon)(3-\epsilon) \frac{\Gamma(\epsilon-2)}{1} \Delta^{'2-\epsilon}_{ij}. 
\end{align}
and $I(2,1)$ as shown in Eq.~\eqref{eq:I21}.

Even though a pole $\frac{2\pi}{\alpha' k_1 \cdot k_2}$ arises from pinching $z_1$ and $z_2$, it is always canceled by the $k_1\cdot k_2$ induced by OPE or $k_1\cdot k_2$ contained in $\Delta'_{12}$. Finally, the loop correction from the pinched diagram $|z_{12}\sim \frac{1}{\tau_2}|$ is proportional to 
\begin{equation}
	M_{3(1)}^{\text{P}}(|z_{12}|\sim \frac{1}{\tau_2}) \sim k_1\cdot k_2 \log (\alpha 'k_1\cdot k_2).
\end{equation}
The same arguments as those for the non-pinched diagrams show that 
this is not a renormalization of the gravitational constant but rather a higher-dimensional operator.
The same reasoning applies to the other pinched diagrams.

In conclusion, perturbative gravitational renormalization is absent in any string model obtained by compactifying the heterotic string,
\begin{equation}
	\beta_{\text{gravity}}=0.
\end{equation}
However, it remains unclear whether quantum field effects can modify the gravitational coupling through quantum corrections to the dilaton vacuum once more realistic moduli stabilization \cite{Grana:2005jc,Douglas:2006es} is taken into account. Such effects cannot be computed within perturbative string theory, highlighting the need for non-perturbative approaches.

\subsection{The gravitational correction to the gauge beta function}
\label{sec:gravi-gauge}
Note that in Section \ref{sec:gauge-beta}, we did not consider $I_L^{\text{vector}}$ and $I_L^{\text{scalar}}$. According to Table~\ref{table:double-copy}, these two terms represent corrections from the 10-dimensional gravitational sector, including the 4-dimensional gravitational sector and other extra-dimensional components, such as graviphotons. Since $W_0=0$,
\begin{align}
I_L^{\text{vector}}&=\sum_{\mu=1}^d T_\mu^{(2)}|_{\text{multi-linear}}, \\
I_L^{\text{scalar}}&=\sum_{\mu=d}^{n_s+d} T_\mu^{(2)}|_{\text{multi-linear}}.
\end{align}  
When the external states are four-dimensional massless gluons without extra-dimensional momentum or polarization vectors, it is straightforward to obtain
\begin{equation}
	I_L^{\text{scalar}}=0.
\end{equation}
Thus we arrive at the first conclusion: there is no correction to the gauge beta function from other extra-dimensional components of higher-dimensional gravity. 

Equations~\eqref{eq:gzq2}, \eqref{eq:p1gzq2}, and \eqref{eq:p2gzq2} show that $\partial G(z)$ and $\partial^2 G(z)$ do not contribute; only $G(z)$ contributes. Hence the nontrivial part of $I_L^{\text{vector}}$ comes only from the $q$-expansion of $\mathcal{J}_3$. One immediately finds that 

\begin{enumerate}
\item For the non-pinched diagram, momentum conservation and the on-shell conditions lead to
\begin{equation}
	I_L^{\text{vector}}\sim \sum_{i<j} k_i \cdot k_j =0.
\end{equation}  
\item For the pinched diagram in which $z_{ij} \rightarrow 0$,
\begin{equation}
	\text{Res}_{z_{ij}\rightarrow 0} I_L^{\text{vector}}\sim -k_i \cdot k_j J_3^{(ij)}.
\end{equation}
\end{enumerate}

This reveals that the gravitational sector, including the graviton, antisymmetric tensor, and dilaton, does provide nonzero quantum corrections to the three-gluon one-loop diagram when some supersymmetry is broken. However, as argued in Section~\ref{sec:gravi-beta} regarding the renormalization conditions for the gravitational beta function, this loop correction generates higher-dimensional operators rather than renormalizing the gauge coupling. Thus, we always have 
\begin{equation}
	\Delta_{\text{gravity}}\beta_{\text{gauge}}=0.
\end{equation}

Using the internal-loop decomposition and double-copy framework, our analysis verifies these results across a broad class of theories derived from heterotic string theory, in contrast to case-by-case Feynman-diagram computations. This result also relies on a perturbative approach, and the non-perturbative correction from gravity to the gauge coupling remains unknown.

\section{\hspace*{-2.5mm}Summary}
\label{sec:summary}
In this paper, the double-copy-like decomposition for the internal loop states, as introduced in Refs.~\cite{Tourkine:2012vx,Geyer:2015jch}, was developed and generalized to bosonic (Section \ref{sec:bstring}) and heterotic (Section \ref{sec:H-string}) string theories in Section \ref{sec:ILDC}. Compared to the formulation in Refs.~\cite{Tourkine:2012vx,Geyer:2015jch}, we employed the chiral-splitting formalism to make the double-copy structure manifest, and included the chiral Koba-Nielsen factors $\mathcal{J}_n$ and $\bar{\mathcal{J}}_n$ in the $q\bar{q}$-expansion. These factors are not essential for superstring theories or the supersymmetric right movers of the heterotic string, since $\mathcal{J}_n$ and $\bar{\mathcal{J}}_n$ contribute non-trivially only for $q^{n \geq 2}$ and $\bar{q}^{n \geq 2}$, whose partition functions exhibit poles of the form $\sim \{1/q, 1/\bar{q}\}$. However, the bosonic string partition function contains poles $\sim \{1/q^2, 1/\bar{q}^2\}$, and the left movers of the heterotic string feature poles $\sim \{1/q^2, 1/\bar{q}^2, 1/q, 1/\bar{q}\}$, making the chiral Koba-Nielsen factors non-negligible. Furthermore, the fermionic realization of the color degrees of freedom in the heterotic string was emphasized, and the corresponding decomposition was formulated in Section \ref{sec:H-string}.

In Section \ref{sec:app-beta}, we applied internal-loop-state decomposition and the double copy to compute the physical beta functions for the gauge coupling constant and the gravitational coupling constant. The detailed scattering integrands were presented in Section \ref{sec:3pt-1loop}. Thanks to the chiral-splitting formalism, the double-copy structure of the integrands became manifest. One can observe that the difference between these two integrands lies in their left-moving parts, which are $\Phi_{\alpha_L\beta_L}$ and $\Lambda^b$, respectively. In the quantum field theory computation of the beta functions, multiple Feynman diagrams contribute. These diagrams are realized in the string framework through different regions of moduli space, and the corresponding decomposition was discussed in Section \ref{sec:string-d-moduli}.

The general QCD beta function was reproduced in Section \ref{sec:gauge-beta}. Unlike previous approaches that extract gauge beta functions from string amplitudes using specific models, our method does not rely on any specialized model and directly yields the general result. In addition, we explored the possible gravitational beta function in Section \ref{sec:gravi-beta}. It was found that $\Lambda^b$ contains a more intricate dependence on the loop momentum $l$, leading to nontrivial Mandelstam factors $s_{ij}$ in the final expression. A careful analysis of renormalization revealed that these quantum corrections involving additional Mandelstam factors should be interpreted as higher-dimensional operators rather than as renormalizations of the original gravitational coupling constant.

Finally, we studied the gravitational correction to the three-gluon one-loop amplitude in Section \ref{sec:gravi-gauge}. This correction arises solely from the chiral Koba-Nielsen factors $\mathcal{J}_3$ and is always accompanied by Mandelstam variables. As a result, it cancels due to momentum conservation in non-pinched diagrams, but remains nonzero in pinched diagrams. However, a similar renormalization analysis indicates that this contribution should also be interpreted as a higher-dimensional operator, not as a renormalization of the gauge coupling constant.

\acknowledgments
I sincerely thank Hong-Jian He for suggesting and discussing the problem of loop corrections to gauge and gravitational beta functions. The two figures in this paper are adapted from an unpublished draft with Hong-Jian He on string loop corrections to gauge beta functions using the traditional RNS formalism without internal-line decomposition and the chiral-splitting formalism.

\newpage

\appendix 

\section{Spin-structure summation}
\label{sec:spin-sum}

There are $2^{2g}$ different spin structures for a genus-$g$ Riemann surface. The final physical result does not depend on the spin structure, so this is usually regarded as a drawback of the RNS formalism. The GS and pure-spinor formalisms can yield the same result without spin-structure summation.

We recall that the world-sheet fermion two-point correlation function for a given spin structure $v$ is given by
\begin{equation}
	\langle\psi^M(z) \psi^N(0)\rangle_v \sim \eta^{MN}  \partial G(z), v=1;  \eta^{MN}  S_v(z), v=2,3,4 . 
\end{equation}
Here we define
\beqs 
\begin{align}
S_v(z) & = \frac{\theta_1'(0) \theta_v(z)}{\theta_v(0) \theta_1(z)},v=2,3,4 
\\
G(z,\bar z)&= -\frac{1}{2} \ln |\frac{\theta_1(z,\tau)}{\theta'_1(0,\tau)}|^2 + \frac{(\operatorname{Im}z)^2 }{4\pi\tau_2}
\end{align}
\eeqs 
We then consider the following Kronecker--Eisenstein series and the associated doubly periodic function,
\begin{align}
	F(z,\eta,\tau) &= \frac{\theta_1'(0) \theta_1(z+\eta)}{\theta_1(\eta) \theta_1(z)}, \\
	\Omega(z,\eta,\tau) &=e^{2\pi i \eta \frac{Im z}{\tau_2}} F(z,\eta,\tau).
\end{align}
It is straightforward to see that 
\begin{align}
	&F(z_1-z_2,\eta,\tau)F(z_2-z_3,\eta,\tau)...F(z_n-z_1,\eta,\tau) \nonumber \\
	=& \Omega(z_1-z_2,\eta,\tau)\Omega(z_2-z_3,\eta,\tau)...\Omega(z_n-z_1,\eta,\tau) \nonumber \\
	=& \sum_n \eta^n V_n(z_1,...,z_n,\tau).
\end{align}
Here $V_n(z_1,...,z_n,\tau)$ is a function of $f_i(z,\tau)$.

The theta functions can be mapped to each other through the following relations,
\begin{align}
	\theta_2(z+\frac{1}{2},\tau) & = -\theta_1(z,\tau), \\
	\theta_4(z+\frac{\tau}{2},\tau) & = ie^{-i\pi z} q^{-\frac{1}{8}}\theta_1(z,\tau), \\
	\theta_4(z+\frac{1}{2},\tau) & = \theta_3(z,\tau), \\
	\theta_3(z+\frac{\tau}{2},\tau) &=e^{-i\pi z} q^{-\frac{1}{8}}\theta_2(z,\tau). 
\end{align}
Thus we define
\begin{align}
	w_2&=\frac{1}{2}, \\
	w_3&=-\frac{1+\tau}{2}, \\
	w_4&=\frac{\tau}{2}
\end{align}
and 
\begin{align}
	\Omega(\vec z_n,w_v,\tau)=&\sum_v S_v(z_1-z_2,\tau)S_v(z_2-z_3,\tau)...S_v(z_n-z_1,\tau)\nonumber \\
	=& \sum_v F(z_1-z_2,w_v,\tau)F(z_2-z_3,w_v,\tau)...F(z_n-z_1,w_v,\tau) \nonumber \\
	=&\sum_v\Omega(z_1-z_2,w_v,\tau)\Omega(z_2-z_3,w_v,\tau)...\Omega(z_n-z_1,w_v,\tau). 
\end{align}

Finally, we evaluate
\begin{equation}
	\mathcal G_n(z_1,...,z_n,\tau)=\sum_v (-1)^{v+1} (\frac{\theta_v(0,\tau)}{\theta'_1(0,\tau)})^4 S_v(z_1-z_2,\tau)S_v(z_2-z_3,\tau)...S_v(z_n-z_1,\tau).
\end{equation}
Using  
\begin{align}
	(-1)^{v+1} (\frac{\theta_v(0,\tau)}{\theta'_1(0,\tau)})^4=&\frac{1}{(e_1-e_2)(e_1-e_3)}, v=2 \\
	(-1)^{v+1} (\frac{\theta_v(0,\tau)}{\theta'_1(0,\tau)})^4=&\frac{1}{(e_2-e_1)(e_2-e_3)}, v=3 \\
	(-1)^{v+1} (\frac{\theta_v(0,\tau)}{\theta'_1(0,\tau)})^4=&\frac{1}{(e_3-e_1)(e_3-e_2)}, v=4 
\end{align}
where $e_v=\wp(w_v)$. Moreover,
\begin{equation} \label{eq:spin-sum-V3-V7}
	\Omega(\vec z_n,\eta,\tau)=\sum_k^{n-2} \frac{(-1)^{n-k}}{(n-k-1)!} (\wp^{(n-k-2)}(\eta)-\hat G_{n-k-2}) V_k +V_n.
\end{equation}
In addition, $\partial \wp(z)$ vanishes at $z=w_v$. Thus $\mathcal G_n$ can generally be expressed in terms of $V_k$ and $e_v$.
Some useful results in the literature are
\beqs 
\label{eq:spin-sum-result}
\begin{align}
	&\mathcal G_{n\leqq 3}=0 \,, \\
	&\mathcal G_{4\leqq n\leqq7}=V_n \,.
\end{align}
\eeqs 

It was noticed in Ref. \cite{Berg:2016wux,Berg:2016fui} that
\begin{equation}
	\prod_i \frac{\theta \Big[ \begin{aligned} \alpha + a_i \\ \beta+b_i \end{aligned} \Big] (0,\tau) }{\theta \Big[ \begin{aligned}
			1/2 + a_i \\ 1/2+b_i \end{aligned} \Big ] (0,\tau)} = \prod_i \frac{\theta \Big[ \begin{aligned} \alpha  \\ \beta \end{aligned} \Big] (b_i + a_i \tau,\tau) }{\theta \Big[ \begin{aligned}
			1/2  \\ 1/2 \end{aligned} \Big ] (b_i + a_i \tau,\tau)}.
\end{equation}
{\bf Here we used the fact that $\sum_i a_i=0$, so this trick works only for supersymmetric string theories.}
Thus we expect that  
\begin{align}
	&\sum_{\alpha,\beta} \eta_{\alpha,\beta} \frac{\theta \Big[\begin{aligned}
			\alpha \\ \beta
		\end{aligned}\Big](0,\tau) \theta \Big[\begin{aligned}
			\alpha+a_1 \\ \beta+b_1
		\end{aligned}\Big](0,\tau) \theta \Big[\begin{aligned}
			\alpha+a_2 \\ \beta+b_2
		\end{aligned}\Big](0,\tau) \theta \Big[\begin{aligned}
			\alpha+a_3 \\ \beta+b_3
		\end{aligned}\Big](0,\tau) }{\theta'\Big[\begin{aligned}
			1/2+a_3 \\ 1/2+b_3
		\end{aligned}\Big](\tau) \theta \Big[\begin{aligned}
			1/2+a_1 \\ 1/2+b_1
		\end{aligned}\Big](\tau) \theta \Big[\begin{aligned}
			1/2+a_2 \\ 1/2+b_2
		\end{aligned}\Big](\tau)\theta \Big[\begin{aligned}
			1/2+a_3 \\ 1/2+b_3
		\end{aligned}\Big](\tau)}  \nonumber \\
	\sim& \sum_{\alpha,\beta} \eta_{\alpha,\beta} \frac{\theta^4 \Big[\begin{aligned}
			\alpha \\ \beta
		\end{aligned}\Big](0,\tau) }{\theta'^4\Big[\begin{aligned}
			1/2+a_3 \\ 1/2+b_3
		\end{aligned}\Big](\tau)} \prod_{i=1}^3 S_{\alpha,\beta}(b_i+a_i\tau,\tau)
\end{align}

The above equation transforms any $n$-point spin-structure summation with orbifold parameters $\{a_i\}$ and $\{b_i\}$ into an $(n+3)$-point spin-structure summation without an orbifold, as long as $\sum_i a_i=0$.

\section{$L$-regularization}
\label{sec:L-regulation}
We consider the following integrals:
\begin{align}
	\label{eq:I1}
	I_1(2n,m)&=\int_{0}^\infty d\tau_2 \int d^{4-\epsilon} l l^{2n} \tau_2^m e^{-\pi\alpha \tau_2 (l^2-\Delta)},\\
	\label{eq:I2}
	I_2(2n,m)&=\int_{L}^\infty d\tau_2 \int d^{4} l l^{2n} \tau_2^m e^{-\pi\alpha \tau_2 (l^2-\Delta)}. 
\end{align}
We first consider the integral over loop momentum. Setting $n=0$ for simplicity,
\begin{align}
	\int d^{d} l e^{-\pi \alpha' \tau_2 l^2} = (\pi \alpha' \tau_2)^{-\frac{d}{2}}.
\end{align}
For $n\neq 0$, one finds the following recursive relation:
\begin{align}
	\int d^{d}l  l^{2n}e^{-\pi \alpha' \tau_2 l^2} = & \int d^{d}l  (-\frac{1}{\pi \alpha'}\frac{\partial}{\partial \tau_2})^ne^{-\pi \alpha' \tau_2 l^2} \nn \\
	=& (-1)^n \frac{\Gamma(\frac{d}{2}+n+1)}{\Gamma(\frac{d}{2})} (\pi \alpha'\tau_2)^{-\frac{d}{2}-n}.
\end{align}

Thus $I_1(n,m)$ and $I_2(n,m)$ can be rewritten as 
\begin{align}
	\label{eq:I1-p}
	I^1(2n,m)&=(-1)^n  \frac{\Gamma(n+3-\epsilon)}{(\pi \alpha')^{n+2-\epsilon}\Gamma(2-\epsilon)}\int_{0}^\infty d\tau_2 \tau_2^{m-n-2+\epsilon} e^{\pi\alpha \tau_2 \Delta},\\
	\label{eq:I2-p}
	I^2(2n,m)&=(-1)^n  \frac{\Gamma(n+3)}{(\pi \alpha')^{n+2}\Gamma(2)}\int_{L}^\infty d\tau_2 \tau_2^{m-n-2} e^{\pi\alpha \tau_2 \Delta}, 
\end{align}
where we assume $\Delta <0$. 
A direct calculation with Mathematica shows that 
\begin{align}
	\int_{0}^\infty d\tau_2 \tau_2^{-n'+\epsilon} e^{\pi\alpha \tau_2 \Delta}&=(\pi \Delta)^{n'-1}(\int d\alpha')^{n'-1} \Gamma(\epsilon)(\pi\alpha' \Delta)^{-\epsilon} \nn  \\
	&= (\pi \Delta)^{n'-1}(\int d\alpha')^{n'-1} (\frac{1}{\epsilon}-\gamma -\log{\pi
		\alpha' \Delta} +\mathcal{O}(e)),  \\  
	\int_{L}^\infty d\tau_2 \tau_2^{-n'} e^{\pi\alpha \tau_2 \Delta} &= (\pi \Delta)^{n'-1}(\int d\alpha')^{n'-1} (-\log(L)-\gamma -\log{\pi
		\alpha' \Delta} +\mathcal{O}(L)).  \ 
\end{align}
This displays the formal mapping $\frac{1}{\epsilon} \leftrightarrow -\log(L)$ and $\epsilon \leftrightarrow L$. However, because there is also $\epsilon$ dependence in $\frac{\Gamma(n+3-\epsilon)}{(\pi \alpha')^{n+2-\epsilon}\Gamma(2-\epsilon)}$ in Eq.~\eqref{eq:I1-p}, 
we have 
\begin{equation}
	I^1(2n,m)-I^2(2n,m)=(\pi \alpha' \Delta)^{m-n-3}\times \text{Const}.
\end{equation}

Any constant can be absorbed into the logarithm function with
\begin{equation}
	\log(\pi\alpha'\Delta) + \log(C) =\log(C\pi\alpha'\Delta).
\end{equation}
In this sense, $I^1(2n,m)$ and $I^2(2n,m)$ are physically equivalent regularization schemes.

\newpage
\bibliographystyle{unsrt}
\bibliography{ILDC-beta}

\end{document}